\documentclass[11pt]{article}

\usepackage{amssymb,amsmath,amsfonts}
\usepackage{graphicx}
\usepackage{color}
\usepackage{ulem}             % \sout
\usepackage{natbib}
\usepackage{tikz}
\def\norm#1{\left\Vert#1\right\Vert}

\def\Frac#1#2{{{\displaystyle\strut#1}\over{\displaystyle\strut#2}}}
\def\Dron#1#2{\Frac{\partial#1}{\partial#2}}
\def\Der#1#2{\Frac{d#1}{d#2}}

\def\C{\mathbb{C}}
\def\R{\mathbb{R}}

\def\T{\mathbb{T}}
\def\N{\mathbb{N}}
\def\Sup{\mathop{\rm Sup}\nolimits}

\newcommand{\figpath}{.}

\newcommand\lam{{\lambda} }
\newcommand\Lam{{\Lambda} }
\newcommand\sig{{\sigma} }

\newcommand\eps{{\varepsilon} }

\newcommand\br{{\bf r} }

\newcommand\brt{\tilde{\bf r} }

\newcommand\xb{{\overline x} }

\newcommand\Bb{{\overline B} }
\newcommand\Hb{{\overline{\! H} }}

\newcommand{\qtext}[1]{\quad \text{#1}\quad}

\newcommand\cG{{\cal G} }
\newcommand\gO{{\cal O} }
\newcommand\cS{{\cal S} }

\newcommand\cC{{\cal C} }
\newcommand\cH{{\cal H} }
\newcommand\cL{{\cal L} }

\newcommand{\md}[1]{{\text{\d{$#1$}}}}
\newcommand\zetas{ \md{\zeta{}} }
\newcommand\Zs{ \md{Z{}} }

\newcommand\xs{ \md{x{}} }
\newcommand\xbs{ \md{\xb{}} }

\newcommand\bfp{{\bf p} }
\newcommand\bfq{{\bf q} }

\newcommand{\be}{\begin{equation}}
\newcommand{\ee}{\end{equation}}

\newcommand{\bpm}{\begin{pmatrix}}
\newcommand{\epm}{\end{pmatrix}}

\newenvironment{disarray}%
 {\everymath{\displaystyle\everymath{}}\array}%
 {\endarray}

\newtheorem{theorem}{Theorem}
\newtheorem{remark}{Remark}

\RequirePackage{fix-cm}
\usepackage{graphicx}

\begin{document}

\title{Rigorous treatment of the averaging process for co-orbital motions in the planetary problem}
%\titlerunning{Averaging process for co-orbital motions}
\author{Philippe Robutel\footnote{IMCCE, Observatoire de Paris, UPMC, CNRS UMR8028, 77 Av. Denfert-Rochereau, 75014 Paris, France} Laurent Niederman, Alexandre Pousse.}

\author{Philippe Robutel\footnote{IMCCE, Observatoire de Paris, UPMC, CNRS UMR8028, 77 Av. Denfert-Rochereau, 75014 Paris, France} \and Laurent Niederman\footnote{Universit\'e Paris XI, LMO, \'equipe Topologie et Dynamique, B\^atiment 425, 91405 Orsay, France}$\;\, ^{,*}$ \and Alexandre Pousse$^{*}$}

%\institute{
%P. Robutel \at
%              IMCCE, Observatoire de Paris, UPMC, CNRS UMR8028, 77 Av. Denfert-Rochereau, 75014 Paris, France \\
%               \email{Philippe.Robutel@obspm.fr}           %  \\
%%             \emph{Present address:} of F. Author  %  if needed
%\and
%A. Pousse \at
%              IMCCE, Observatoire de Paris, UPMC, CNRS UMR8028, 77 Av. Denfert-Rochereau, 75014 Paris, France \\
%               \email{Alexandre.Pousse@obspm.fr}           %  \\
%L. Niederman \at
%              Universit\'e Paris XI, LMO, \'equipe Topologie et Dynamique, B\^atiment 425, 91405 Orsay, France \\
%              and \\
%              IMCCE, Observatoire de Paris, UPMC, CNRS UMR8028, 77 Av. Denfert-Rochereau, 75014 Paris, France \\
%              \email{Laurent.Niederman@math.u-psud.fr}
%}

\date{\today}


\maketitle

\begin{abstract}
We develop a rigorous analytical Hamiltonian formalism adapted to the study of the motion of two planets in co-orbital resonance.
By constructing a complex domain of holomorphy for the planetary Hamiltonian, we estimate the size of the transformation that maps this Hamiltonian to its first order averaged over one of the fast angles.
After having derived an integrable approximation of the averaged problem, we bound the distance between this integrable approximation and the averaged Hamiltonian. This finally allows to prove rigorous theorems on the behavior of co-orbital motions over a finite but large timescale.

 \bigskip

%\keywords{Co-orbital resonance \and Lagrange configurations \and Euler configuration \and Three-body problem}
 %\subclass{70F10 \and 70F15 \and 70H08 \and 70H09 }

\end{abstract}

\section{Introduction}
\label{sec:intro}

Averaging methods are common techniques to study the dynamics of Hamiltonian systems in celestial mechanics. The first and most famous example of averaged Hamiltonian system is perhaps the secular planetary problem, where the Hamiltonian of the planetary problem is averaged over the mean longitudes of the planets.   The secular equations of the planetary motion appear in Lagrange's work on stability of the solar system \citep{Lagrange1778} while the secular Hamiltonian appears in Delaunay's memory about the theory of the Moon \citep{Delaunay1860}.    \cite{Poi1892} gave an expression of the secular Hamiltonian of the planetary three body problem while \cite{LaRo1995} present an analytical  method allowing to compute the expansion of the planetary Hamiltonian, and in particular, to get a concise expression of its secular part.

 The secular Hamiltonian of the planetary problem can be obtained, up to a finite order of the planetary masses, by averaging over the planetary mean longitudes (construction of a resonant normal form), the symplectic transformation that maps the non-averaged Hamiltonian to the secular one being close to the identity. Precise estimates on the size of these transformations are required in order to prove the existence of invariant tori using the KAM theory.   These kinds of rigorous estimates have been established especially by  \cite{Arnold1963}, \cite{Fejoz2004} and \cite{ChPi2011}.

When two planets are in co-orbital resonance, and more generally for two planets in mean-motion resonance, the transformation leading to the secular Hamiltonian is no more close to the identity, even at first order. Consequently, the secular motion does not provide a good representation of the real planetary motion.
In these cases, it is still possible to use averaging methods, but for two planets the Hamiltonian is generally averaged over one fast angle, that is one of the planetary mean longitudes.

Many authors work with the resonant averaged problem.
 Some of them use  analytic approximations of the averaged Hamiltonian or averaged motion \citep[e.g.][]{Ed1977, Namouni1999, Morais2001,RoPo2013}, while others prefer a numerical averaging \citep[e.g.][]{NeThoFeMo02, GiuBeMiFe2010}.
But in none of these works the size of the transformation and consequently the "distance" between the secular and the "complete" solution is rigorously estimated.

In this paper, we estimate this distance and give an upper bound of the time for which this difference remains small enough.
For this purpose,
we derive, in section \ref{sec:hamperturbationtheo},  a complex domain of holomorphy for the planetary Hamiltonian which allows to compute quantitatively the size of the transformations and of the perturbations involved in our construction.

  All the computations about perturbation methods are derived in section \ref{seq:pert_theory} and constitute the main novelty of this paper.

  The topology of the averaged Hamiltonian is studied in section \ref{sec:averadgedro} where we show the existence of an invariant manifold,  associated to the quasi-circular motions, which carry an integrable dynamic.

  Section \ref{sec:simplified} is devoted to the construction of an integrable approximation of the averaged Hamiltonian and its degree of accuracy in the vicinity of the invariant manifold considered in section \ref{sec:averadgedro}. We also give the general form of the solutions of this integrable system.

  Finally, in section \ref{se:cons_coorb}, we can combine the quantitative estimates given in section \ref{seq:pert_theory} and the bounds on the remainders between the averaged Hamiltonian and its integrable approximation. This allows to prove rigorous theorems on the behaviour of co-orbital motions over large timescales.

   The last section concerns the proof of the technical propositions and lemma used in our reasonings.

 \section{Hamiltonian setting of the problem}
\label{sec:hamperturbationtheo}
\subsection{Canonical heliocentric coordinates}
We consider two planets of respective masses ${\widetilde{m}}_1$ and ${\widetilde{m}}_2$ orbiting a central body (Sun, or
star) of mass $m_0$ dominant with respect to the planetary masses.
As only co-orbital planets are considered, no planet is permanently farther from the central body than the other, so the
heliocentric coordinate system seems to be the most adapted to this situation. Following \cite{LaRo1995}, the Hamiltonian
of the three-body problem reads

\be
\begin{split}
{\widetilde{\mathcal{H}}}({{\widetilde{\bf R}}_j},\br_j) ={\widetilde{\mathcal{H}}}_K({{\widetilde{\bf R}}_j,\br_j}) + {\widetilde{\mathcal{H}}}_P({{\widetilde{\bf R}}_j},\br_j) \quad \text{with}  \\
{\widetilde{\mathcal{H}}}_K({{\widetilde{\bf R}}_j},\br_j) =  \sum _{j\in\, \{1,2\}} \left(
 \frac{{{\widetilde{\bf R}}_j}^2}{2{{\widetilde{\beta}}_j}} -   \frac{\widetilde{\mu}_j}{\norm{\br_j}}
    \right)
    \quad \text{and} \\
{\widetilde{\mathcal{H}}}_P({\widetilde{\bf R}}_j,\br_j) = \frac{{\widetilde{\bf R}}_1\cdot{\widetilde{\bf R}}_2}{{{m}}_0} - \cG\frac{{\widetilde{m}}_1 {\widetilde{m}}_2}{\norm{\br_1 -\br_2}},
     \end{split}
     \label{eq:ham_cart}
\ee

where $\br_j$ is the heliocentric position of the planet $j$, ${\widetilde{\beta}}_j = m_0{\widetilde{m}}_j(m_0+{\widetilde{m}}_j)^{-1}$ and ${\widetilde{\mu}}_j = \cG(m_0+{\widetilde{m}}_j)$, $\cG$ being the gravitational constant. The conjugated variable of $\br_j$, denoted by ${\widetilde{\bf R}}_j$, is the barycentric linear momentum of the body of index $j$. In this expression, ${\widetilde{\mathcal{H}}}_K$ corresponds to the unperturbed Keplerian motion of the two planets, more precisely the motion of a mass ${\widetilde{\beta}}_j$ around a fixe center of mass $m_0+{\widetilde{m}}_j$, while ${\widetilde{\mathcal{H}}}_P$ models the gravitational perturbations.

 The Hamiltonian ${\widetilde{\mathcal{H}}}$ is analytical on the whole phase space ($\R^{8}$ since we consider the planar problem) except on the manifold which corresponds to collisions between two planets:
$${\widetilde{\mathfrak{D}}}=\left\{ ({\widetilde{\bf R}}_1,\br_1, {\widetilde{\bf R}}_2,\br_2)\in\R^8\ \text{\rm such that}\ \br_1\not=\br_2\right\} .$$

If we introduce the small parameter $\eps$ given by
\be
\eps = \text{Max}\left(\frac{{\widetilde{m}}_1}{m_0},\frac{{\widetilde{m}}_2}{m_0}\right),
\ee
one can verify that the Keplerian term of the planetary Hamiltonian is of order $\eps$ and  that the other one is of order $\eps^2$ as long as the mutual distance between the planets remain large enough. This feature justifies a perturbative approach.

\bigskip

\begin{remark}\label{defconstant}
In the sequel, we will not give explicit estimates of the constants independent of the small parameters involved in the problem in order to avoid cumbersome and meaningless expressions. We will just denote uniformly by $M$ a positive constant chosen sufficiently large for our purpose but independent of the relevant quantities, we will try to specify this at each occurrence of such constant but sometimes it will be omitted.
\end{remark}

\subsection{The rescaled heliocentric coordinates}

According to (\ref{eq:ham_cart}), the momenta ${\widetilde{\bf R}}_j$ and the Keplerian part of the Hamiltonian are of order  $\eps$ while the perturbation is quadratic in $\eps$. In order to get more homogeneous quantities, 
 it is convenient to rescale the planetary  masses, the Hamiltonian and the canonical variables.

  First, we introduce new planetary masses and new reduced masses by the relations: 
$$m_1\! =\!\frac{{\widetilde{m}}_1}{\varepsilon},\ m_2\! =\!\frac{{\widetilde{m}}_2}{\varepsilon}\ ;\ \beta_j\! =\!\frac{{\widetilde{\beta}}_j}{\varepsilon}\! =\!\frac{m_0 m_j}{m_0+\varepsilon m_j}\ \text{\rm and}\ \mu_j\! =\!\cG(m_0+\varepsilon m_j)\ \text{\rm for}\ j\!\in\!\{ 1,2\} .$$
 Moreover, for $j\!\in\!\{ 1,2\}$, we rescale the impulsions by ${\widetilde{\bf R}}_j =\eps\brt_j$ while the positions $\br_j$ remain unchanged. Hence, we have made a conformal symplectic transformation $\mathcal{T}$ which changes the symplectic form with $$\sum_j d\brt_j\wedge d\br_j =\eps^{-1}\sum_j d{\widetilde{\bf R}}_j\wedge d\br_j$$
 and the Hamiltonian linked to the considered system becomes
 $${\mathcal{H}}(\brt_j ,\br_j )=\eps^{-1}{\widetilde{\mathcal{H}}}\circ \mathcal{T}(\brt_j ,\br_j ).$$
 This leads to the expression:
\be
\begin{split}
{\mathcal{H}}(\brt_j,\br_j) = {\mathcal{H}}_K(\brt_j,\br_j) +  \eps {\mathcal{H}}_P(\brt_j,\br_j) \quad \text{with}  \\
 {\mathcal{H}}_K(\brt_j,\br_j) =  \sum _{j\in\, \{1,2\}} \left(
 \frac{\brt_j^2}{2\beta_j} - \frac{\mu_j \beta_j}{\norm{\br_j}}
    \right)
    \quad \text{and} \\
  {\mathcal{H}}_P(\brt_j,\br_j) = \frac{\brt_1\cdot\brt_2}{m_0} -  \cG\frac{m_1m_2}{\norm{\br_1 -\br_2}},
     \end{split}
     \label{eq:ham_cart_renorm}
\ee
the canonical variables $\brt_j$ and $\br_j$ and the Keplerian part ${\mathcal{H}}_K$ being now of order one.

 The three body Hamiltonian ${\mathcal{H}}$ is defined on
$$\mathfrak{D}=\left\{ (\brt_1,\br_1,\brt_2,\br_2)\in\R^8\ \text{\rm such that}\ \br_1\not=\br_2\right\}$$
since the positions remain unchanged.

\medskip

\subsection{The Poincar\'e variables.}

 In order to define a canonical coordinate system related to the elliptic elements $(a_j,e_j,\lam_j,\varpi_j)$ (respectively the semi-major axis, the eccentricity, the mean longitude and the longitude of the pericenter of the planet $j$), we use complex Poincar\'e's rectangular variables $(\lam_j,\Lam_j,x_j,-i\xb_j)_{j\in\{ 1,2\} }\in (\mathbb{T}\times\mathbb{R}\times\mathbb{C}\times\mathbb{C})^2$:
\be\Lam_j = \beta_j\sqrt{\mu_ja_j}
\quad \text{and}\quad x_j = \sqrt{\Lam_j}\sqrt{1-\sqrt{1-e_j^2}}\exp(i\varpi_j),
\ee
this coordinate system has the advantage to be regular when the eccentricities and the inclinations tend to zero.

  Consequently, we have the product of the analytic symplectic transformations around the circular orbits (i.e.  for a given constant $c_0>0$):
\be
\Phi_j\ : \quad \Bigg\{
\begin{array}{ccc}
\mathfrak{E}_j  &\longrightarrow & \C^{4}\\
(\lam_j,\Lam_j,x_j,-i\xb_j )\ \ &\longmapsto  &(\brt_j ,\br_j )
\end{array}
\quad  (j= 1,2)
\ee
with

$$
\mathfrak{E}_j\! =\!\left\{ (\lam_j,\Lam_j,x_j,-i\xb_j )\ {\rm where}\ \left\vert
\begin{array}{rc}
\lam_j &\in\T\ \\
\Lam_j &\in\R^{*+}\ \\
\end{array}\right.
\text{\rm and}\ x_j\in\mathbb{C}\ \text{\rm with}\ \vert x_j\vert\leq c_0\sqrt{\Lam_j }\right\}
$$
which yields the new planetary Hamiltonian:
 $${\widetilde{H}}(\lam_j,\Lam_j,x_j,-i\xb_j)={\widetilde{H}}_K(\Lam_1,\Lam_2) +{\widetilde{H}}_P(\lam_j,\Lam_j,x_j,-i\xb_j).$$

  ${\widetilde{H}}$ is analytic on the domain $\mathfrak{A}\subset (\mathbb{T}\times\mathbb{R}\times\mathbb{C}\times
  \mathbb{C})^2$ defined by:
 $$\left\{(\lam_j,\Lam_j,x_j,-i\xb_j)\in\mathfrak{E}_j\ \text{\rm for}\ j\in\{ 1,2\}\ / \ \left(
 \begin{array}{rc}
 \Phi_1 (\lam_1,\Lam_1,x_1,-i\xb_1)\\
 \Phi_2 (\lam_2,\Lam_2,x_2,-i\xb_2)\\
 \end{array}\right)\in\mathfrak{D}\right\}$$

\subsection{The 1:1 Resonance}
\label{r11}

 The Keplerian part of the Hamiltonian expressed in terms of Poincar\'e's variables reads:
\be
{\mathcal{H}}_K(\brt_j,\br_j)={\widetilde{H}}_K (\Lam_1,\Lam_2)= -\sum _{j\in\, \{1,2\}} \frac{1}{2}\frac{\mu_j^2\beta_j^3}
{\Lam_j^2}
\label{eq:Kep_poinc}
\ee
Consequently, the 1:1 mean motion resonance, also called co-orbital resonance, is reached
when $(\Lam_1,\Lam_2) = (\Lam_1^0,\Lam_2^0)$ such that the two mean motions are the same
(let us denoted by $\omega$ this frequency), that is:
\be
  \frac{\partial{\widetilde{H}}_K}{\partial\Lam_1}(\Lam_1^0,\Lam_2^0)  =  \frac{\partial{\widetilde{H}}_K}
  {\partial\Lam_2}(\Lam_1^0,\Lam_2^0) \, := \omega >0   \, ,
\ee
or:
\be
\frac{\mu_1^2\beta_1^3}{(\Lam_1^0)^3}  =\frac{\mu_2^2\beta_2^3}{(\Lam_2^0)^3}:=\omega >0\, .
\ee
 In order to work in a neighborhood of this resonance, we introduce a new  coordinate system with an affine
 unimodular transformation completed in a symplectic way $\Psi(\zeta_1,\zeta_2, Z_1,Z_2,x_1,x_2)=(\lam_1,\lam_2,\Lam_1,
 \Lam_2,x_1,x_2)$
 with:

\be
\begin{pmatrix}
\zeta_1\\
\zeta_2
\end{pmatrix}
=
\begin{pmatrix}
1 & -1\\
0 & 1
\end{pmatrix}
\begin{pmatrix}
 \lam_1 \\
  \lam_2
\end{pmatrix}
, \quad
\begin{pmatrix}
Z_1\\
Z_2
\end{pmatrix}
=
\begin{pmatrix}
1 & &0\\
1 & &1
\end{pmatrix}
\begin{pmatrix}
 \Lam_1  - \Lam_1^0 \\
 \Lam_2  - \Lam_2^0
\end{pmatrix}
 \label{eq:transf_lin}
 \ee
which yields slow-fast angles.

More precisely, in a neighborhood of the co-orbital resonance,  the angular variables evolve at different
rates: $\zeta_2$ is a "fast" angle with a frequency of order $1$,  $\zeta_1$ undergoes "semi-fast" variations at a frequency of order $\sqrt\eps$ (see Section \ref{sec:sol_C_0}), while the variables $x_j$ related to the eccentricities are associated to the ÓslowÓ degrees of freedom evolving on a time scale of order $\eps$ (secular variations).
  With this set of variables, the planetary Hamiltonian becomes:
 \be
 \begin{split}
 H(\zeta_j, Z_j, x_j, -i\xb_j)\!  &=\! {\widetilde{H}}\circ\Psi (\zeta_j, Z_j, x_j, -i\xb_j) \\
                                               &=\! H_K(Z_1, Z_2) +H_P(\zeta_j, Z_j, x_j, -i\xb_j) .
\end{split}
\ee
\bigskip

 For an arbitrary $\Delta >0$, we consider the following domain centred at the 1:1 keplerian resonance defined in the
 $(\zeta, Z, x, -i\xb)$ variables:
$$\mathcal{K}^{(0)}_\Delta\! =\!\left\{\left( (\zeta_1,0,0,0),(\zeta_2,0,0,0)\right)\!\in\! (\mathbb{T}\times\mathbb{R}
\times\mathbb{C}\times\mathbb{C})^2\ \text{\rm such that}\ \vert\zeta_1\vert >\Delta\right\}$$
where $\vert .\vert$ denotes the usual distance over the quotient space $\mathbb{T}=\mathbb{R}/2\pi\mathbb{Z}$.
 Hence, the angular separation $\Delta$ over $\mathcal{K}^{(0)}_\Delta$ yields a minimal distance between the planets:
% $$
% \delta\! :=\! 2a\sin\left(\frac{\Delta}{2}\right)\ \text{\rm where}\ a\! :=\!\min\!\left(\!\frac{(\Lambda_1^0)^2}
% {\beta_1^2\mu_1},\frac{(\Lambda_2^0)^2}
% {\beta_2^2\mu_2}\right)\ \text{\rm is the lowest radius of the
% orbits.}
% $$

%
\begin{equation*}
  \delta\! :=\! 2a\sin\left(\frac{\Delta}{2}\right) \qtext{\rm where}
 a\! :=\!\min\!\left(\!\frac{(\Lambda_1^0)^2}
 {\beta_1^2\mu_1},\frac{(\Lambda_2^0)^2}
 {\beta_2^2\mu_2}\right)
 \end{equation*}
is the lowest radius of the orbits.
\medskip

 The condition $\vert\zeta_1\vert >\Delta$ for $\zeta_1\in\mathbb{T}$ can also be considered with the
 real variable $\zeta_1\in ]\Delta , 2\pi -\Delta [$ since there exists an unique real representative
 in this interval for an angle with a modulus lowered by $\Delta$. Hence $\mathcal{K}^{(0)}_\Delta$ has
 the structure of a cylinder in $(]\Delta , 2\pi -\Delta [\times\mathbb{R}\times\mathbb{C}
 \times\mathbb{C})\times (\mathbb{T}\times\mathbb{R}\times\mathbb{C}\times\mathbb{C})$.

 For $\rho >0$ and $\sigma >0$ small enough (this will be specified in the sequel), our initial domain $\mathcal{K}^{(0)}_\Delta$ can be extended in a complex neighbourhood
 $\mathcal{K}^{(\mathbb{C})}_{\Delta ,\rho ,\sig}\subset\mathbb{C}^8$ of the following type:
%
% $$
% \mathcal{K}^{(\mathbb{C})}_{\Delta ,\rho ,\sig}\! =\!\left\{\left(\zeta_j,Z_j,\xi_j ,\eta_j\right)_{j\in\{ 1,2\} }\!
% \in\!\mathbb{C}^8\ /\
% \left(\begin{array}{rc}
%({\text{\rm Re}}(\zeta_1),0,0,0)\\
% ({\text{\rm Re}}(\zeta_2),0,0,0)\\
% \end{array}
% \right)\!\in\!\mathcal{K}^{(0)}_\Delta\
% \text{\rm and}\
%\begin{array}{cl}
% \diamond &\vert Z_j\vert\leq\rho  \\
% \diamond &\vert\xi_j\vert\leq\sqrt{\rho\sigma} \\
% \diamond &\vert\eta_j\vert\leq\sqrt{\rho\sigma} \\
% \diamond &\vert\text{\rm Im}(\zeta_j)\vert\leq\sigma
%\end{array}
%\!\!\!
% \right\}
% $$
%
 \begin{equation*}
\begin{split}
 \mathcal{K}^{(\mathbb{C})}_{\Delta ,\rho ,\sig}\! = \bigg\{ & \left(\zeta_j,Z_j,\xi_j ,\eta_j\right)_{j\in\{ 1,2\} }\!
 \in\!\mathbb{C}^8\ /\
 \left(\begin{array}{rc}
({\text{\rm Re}}(\zeta_1),0,0,0)\\
 ({\text{\rm Re}}(\zeta_2),0,0,0)\\
 \end{array}
 \right)\!\in\!\mathcal{K}^{(0)}_\Delta\
 \\
&  \qtext{\rm and}
\vert Z_j\vert\leq\rho,\,
\vert\xi_j\vert\leq\sqrt{\rho\sigma} ,\,
\vert\eta_j\vert\leq\sqrt{\rho\sigma} ,\,
\vert\text{\rm Im}(\zeta_j)\vert\leq\sigma
\bigg\}
\end{split}
 \end{equation*}
actually
$$
{\mathcal{K}^{(0)}_\Delta}\subset{\mathcal{K}^{(\mathbb{C})}_{\Delta ,\rho ,\sig}}\cap{\mathfrak{A}}=\!
\left\{\left(\zeta_j,Z_j,\xi_j ,\eta_j\right)_{j\in\{ 1,2\} }\!\in\mathcal{K}^{(\mathbb{C})}_{\Delta ,\rho ,\sig}\
{\text{\rm with}}\ \eta_j =i{\overline{\xi}}_j=x_j\!\right\}.
$$

 In this setting, we can define a complex domain of holomorphy for the planetary Hamiltonian where it will be
possible to estimate the size of the transformations and the functions involved in our construction
of a 1:1 resonant normal form.
 More specifically, for $\Delta >0$, $\rho >0$, $\sigma >0$ and for $p>0$ we will consider the compact:
$$\mathcal{K}_p :=\mathcal{K}^{(\mathbb{C})}_{\Delta ,p\rho ,p\sigma }$$
and the supremum norm $\vert\vert .\vert\vert_{\infty}$ on the space of holomorphic functions over
the compact $\mathcal{K}_p$ which will be denoted $\vert\vert .\vert\vert_p$.

\medskip

  With $H_K$ of class ${\mathcal{C}}^{(3)}$ on the compact $\{ (z_1,z_2)\in\C^2\ /\ \max (\vert z_1\vert ,\vert z_2\vert )\leq\rho\}$, there exists a constant ${\widetilde{\mathrm{M}}}>0$ which satisfies:
% $$
% \forall \, (p_1,p_2) \in \N^2 \,\ \text{such that} \quad \vert p_1\vert + \vert p_2\vert \leq 3 \qtext{ one has } \left\vert\left\vert \partial_{Z_1^{p_1},Z_2^{p_2}}  H_K\right\vert\right\vert_\rho <{\widetilde{\mathrm{M}}}
% $$
\be
 \forall \, (p_1,p_2) \in \N^2 \, \text{such that} \,\, \vert p_1\vert + \vert p_2\vert \leq 3,\, \text{one has:}\, \left\vert\left\vert \partial_{Z_1^{p_1},Z_2^{p_2}}  H_K\right\vert\right\vert_\rho \!<{\widetilde{\mathrm{M}}}
 \label{eq:born_der}
\ee
where we denote $\vert\vert .\vert\vert_\rho$ the supremum norm $\vert\vert .\vert\vert_\infty$ on the space of holomorphic functions over the compact $\{ (z_1,z_2)\in\C^2\ /\ \max (\vert z_1\vert ,\vert z_2\vert )\leq\rho\}$.

By the real-analyticity of the transformation in Poincar\'e resonant action-angle variables $\Upsilon =(\Phi_1\circ\Psi ,\Phi_2\circ\Psi )$, there exist $\rho_0>0$ and $\sigma_0>0$ small enough such that $\Upsilon$ can be extended into a holomorphic function over the compact $\mathcal{K}^{(\mathbb{C})}_{\Delta ,\rho_0 ,\sigma_0}$ with:
 $$
  {\Phi}_j\circ\Psi :  \quad \Bigg\{
 \begin{array}{clc}
\mathcal{K}^{(\mathbb{C})}_{\Delta ,\rho_0 ,\sigma_0}\  &
 \longrightarrow
 & \C^{4} \\
\left(\varsigma_j,z_j,\xi_j ,\eta_j\right)_{ j\in\{ 1,2\} } & \longmapsto &(\brt_j,\br_j)
\end{array}
\quad\text{\rm for}\ j= 1,2
$$
and the differential of $\Psi$ is bounded by a constant $C>0$ with respect to the supremum norm $\vert\vert .
 \vert\vert_{\Delta ,\rho_0,\sigma_0 }$ on the space of holomorphic functions over the compact $\mathcal
 {K}^{(\mathbb{C})}_{\Delta ,\rho_0 ,\sigma_0}$.

 Without loss of generality, we can assume that $C>1$.

\medskip

\begin{theorem}
\label{theo1}

 There exist constants $\rho_0 >0$, $\sigma_0 >0$ and $M\geq {\widetilde{\mathrm{M}}}$ independent of $\varepsilon$ and $\Delta$ (or equivalently of $\delta$) such that if we assume:
 \be
\rho\leq\rho_0,\ \sigma\leq\sigma_0,\ 0<\rho <\sigma <1\ \text{\rm and}\ \sigma\leq \frac{\delta}{16C}=
\frac{a}{8C}\sin\left(\frac{\Delta}{2}\right) <\delta\label{seuil_1}
 \ee
then the following bounds are valid:
$$\frac{\varepsilon}{M} <{\text{\rm Inf}}_{\mathcal{K}_1}\left(\left\vert H_P\left(\varsigma_j,z_j,\xi_j ,\eta_j\right)\right\vert\right)\ {\text{\it and}}\ \vert\vert H_P\vert\vert_1 <M\frac{\varepsilon}{\delta}$$
where we denote $\vert\vert .\vert\vert_1$ the supremum norm $\vert\vert .\vert\vert_\infty$ on the space of holomorphic functions over the compact $\mathcal{K}_1 =\mathcal{K}^{(\mathbb{C})}_{\Delta ,\rho ,\sigma }$.
\end{theorem}

\medskip

{\hskip-0,5truecm\bf{Remark:}} As it was specified, from now on we will denote uniformly by $M$ a positive constant independent of $\varepsilon$, $\rho$, $\sigma$ and $\Delta$ (or equivalently of $\delta$) high enough for our purpose. We will try to specify this at each occurrence of such constant but sometimes it will be omitted

\medskip

\section{Hamiltonian perturbation theory}
\label{seq:pert_theory}

 In this paper, we only consider the averaged Hamiltonian at first order in the planetary masses.
More precisely, we prove quantitatively that there exists a canonical transformation $\cC$
which maps the original Hamiltonian
$H$ in
$$
H'(\zetas_j,\Zs_j, \xs_j, -i\xbs_j)\! =\! H_K(\Zs_1,\Zs_2)\! +\!\Hb_P(\zetas_1,\Zs_j,\xs_j, -i\xbs_j)\! +\!
H_*(\zetas_j,\Zs_j,\xs_j, -i\xbs_j).
$$
In this expression, the Keplerian part $H_K$ reads
\be
H_K(\Zs_1,\Zs_2) =-\frac{\beta_1^3\mu_1^2}{2(\Lam_1^0 + \Zs_1)^2}
                                     -\frac{\beta_2^3\mu_2^2}{2(\Lam_2^0 - \Zs_1 + \Zs_2)^2},
\ee
while the averaged perturbation with respect to the second angle (which corresponds to a time averaging along the
periodic orbits on the torus at the origin for the Kepler problem in our resonant action-angle variables) is given by:
\be
\begin{array}{rl}
\Hb_P(\zetas_1, \Zs_j, \xs_j, -i\xbs_j)\! &=\!\frac{\omega}{2\pi}
{\displaystyle{\int_0^\frac{2\pi}{\omega}}}H_P(\zetas_1,
\zetas_2 +\omega t, \Zs_j, \xs_j, -i\xbs_j) dt\!\\
&=\!\frac{1}{2\pi}{\displaystyle{\int_0^{2\pi}}}H_P(\zetas_1,
\zetas_2 , \Zs_j, \xs_j, -i\xbs_j) d\zetas_2
\end{array}
\ee
The remainder  $H_*(\zetas_j,\Zs_j,\xs_j, -i\xbs_j)$ is a much smaller general perturbation whose size will be estimated in the next section.

\subsection{Hamiltonian perturbation theory}

Now, we specify our construction of the averaging transformation which will be the time-one map $\cC =\Phi_{1}^{\chi}$
of the Hamiltonian flow generated by some auxiliary function $\chi$.

 Using the Poisson bracket:
$$
L_{\chi} (f) =\{ \chi ,f\} = \partial_1\chi .\partial_2 f+\partial_3\chi .
\partial_4 f - \partial_2\chi .\partial_1 f - \partial_4\chi .\partial_3 f,
$$
we have $\cC =\exp ( L_{\chi})$ and $\Hb =\exp ( L_{\chi} ) (H)$.
 In the new variables, the Hamiltonian can be written:
$$
H' =H_K+H_P+ \{ \chi ,H_K\} +\{ \chi ,H_P \} +H' -H -\{ \chi ,H\} .
$$
 With a generating function $\chi$ comparable with $H_P$, the terms of order 1 in this expansion (with respect to $H_P$) are~:
\begin{equation*}
\begin{split}
[H_P+ \{ \chi ,H_K\} ](\zetas_j,\Zs_j,\xs_j,\! -i\xbs_j) &=H_P(\zetas_j,\Zs_j,\xs_j,\! -i\xbs_j) \\
& +\nabla H_K
(\Zs_1,\Zs_2).\frac{\partial\chi}{\partial\zetas}(\zetas_j,\Zs_j,\xs_j,\! -i\xbs_j)
\end{split}
\end{equation*}
 We will choose $\chi$ as a solution, over $\mathcal{K}_1 =\mathcal{K}^{(\mathbb{C})}_{\Delta ,\rho ,\varrho}$, of the equation:
$$
\omega .\frac{\partial\chi}{\partial\zetas_2}(\zetas_j,\Zs_j,\xs_j, -i\xbs_j) =\Hb_P(\zetas_1,\Zs_j,\xs_j, -i\xbs_j)-H_P(\zetas_j,
\Zs_j,\xs_j,-i\xbs_j)\label{equ_homo}
$$
actually, this equation is satisfied by~:
%$$
%\chi (\zetas_j,\Zs_j,\xs_j, -i\xbs_j)=\Frac{\omega}{2\pi}\int_0^{\frac{2\pi}{\omega}}\!\! t.\!\left[ \Hb_P(\zetas_1,\Zs_j,
%\xs_j, -i\xbs_j)-H_P(\zetas_1,\zetas_2 +\omega t,\Zs_j,\xs_j, -i\xbs_j)\right] dt
%$$
\begin{equation*}
\begin{split}
\chi (\zetas_j,\Zs_j,\xs_j, -i\xbs_j)=\Frac{\omega}{2\pi}\int_0^{\frac{2\pi}{\omega}}\!\! t.
&\!\left[
 \Hb_P(\zetas_1,\Zs_j, \xs_j, -i\xbs_j)
 \right. \\
&  \left.
-H_P(\zetas_1,\zetas_2 +\omega t,\Zs_j,\xs_j, -i\xbs_j)
\right] dt
\end{split}
\end{equation*}
and
\begin{equation*}
\begin{split}
&H'(\zetas_j,\Zs_j, \xs_j, -i\xbs_j)  =  \\
&H_K(\Zs_1,\Zs_2)  +\Hb_P(\zetas_1,\Zs_j,\xs_j, -i\xbs_j)  +
H_*(\zetas_j,\Zs_j,\xs_j, -i\xbs_j)
\end{split}
\end{equation*}
$$\text{\rm with}\
H_*\! =\!\left(\nabla H_K\! -\! {\overrightarrow{\omega}}\right).\Dron{\chi}{\zetas}+\{ \chi ,H_P \} +H' -H -\{ \chi ,H\}
\ \text{\rm for}\ {\overrightarrow{\omega}}\! :=\!\left(
                                   \begin{array}{c}
                                     0\\
                                     \omega\\
                                   \end{array}
                                 \right)$$

\medskip

\subsection{Quantitative Hamiltonian perturbation theory}
\label{sec:PerturbTheory}

 Here, we make the construction described in the previous section with accurate estimates on
the size of the Hamiltonian and on the size of the transformations which are involved.
 As in the section $\ref{r11}$, for $\Delta >0$, $\rho >0$ and $\sigma >0$, we consider the compact $\mathcal{K}_2$
and the supremum norm $\vert\vert .\vert\vert_2$ on the space of holomorphic functions over $\mathcal{K}_2$.
 With the expression of $\chi$ and the fact that $\vert\vert\Hb_P\vert\vert_2\leq\vert\vert H_P
\vert\vert_2$, we obtain~:
\be
\vert\vert\chi\vert\vert_2 <{\Frac{2\pi}{\omega}}\vert\vert H_P\vert\vert_2 
\label{seuil_Pert}
\ee
which allows to prove the following:

\medskip

\begin{theorem}
\label{theo2}

 Let $\Delta >0$, $\rho >0$ and $\sigma >0$ such that $(\Delta ,2\rho ,2\sigma )$ satisfy the condition (\ref{seuil_1}), hence:
\be
\rho <\!\frac{\rho_0}{2} ;\ \rho <\!\sigma <\!\min\!\left(\!\frac{\sigma_0}{2},\frac{\delta}{32C}\!\right)\ \text{\rm for}\
\delta\! :=\!\sin\left(\!\frac{\Delta}{2}\!\right)\min\!\left(\!\frac{(\Lambda_1^0)^2}{\beta_1^2\mu_1},\frac
{(\Lambda_2^0)^2}{\beta_2^2\mu_2}\right) .\label{seuil_2}
 \ee
 In order to get a small enough transformation (or equivalently a small enough $\chi$), we assume moreover that
\be
\frac{\varepsilon}{\delta\rho\sigma}<\frac{\omega}{32\pi M} .\label{seuil_3}
\ee
 Then there exists a canonical transformation ${\cal C} : {\cal K}_{\!\frac32}\rightarrow
{\cal K}_2$ such that
\be
{\cal C}\ {\rm is\ one\! -\! to\! -\! one\ and}\ {\cal K}_{\!\frac54}
\subseteq{\cal C}({\cal K}_{\!\frac32})\subseteq{\cal K}_{\!\frac74}\label{taille_transfo}
\ee
and, still using the notations
$$(\zeta_j,Z_j,x_j, -i\xb_j)_{ j\in\{ 1,2\} }={\cal C}\left(\zetas_j,\Zs_j,
\xs_j, -i\xbs_j\right)_{ j\in\{ 1,2\} },
$$
the transformed Hamiltonian $H'=H\circ{\cal C}$ can be written as
$$H'(\zetas_j,\Zs_j, \xs_j, -i\xbs_j)\! =\! H_K(\Zs_1,\Zs_2)\! +\Hb_P(\zetas_1,\Zs_j,\xs_j, -i\xbs_j)\! +\!
H_*(\zetas_j,\Zs_j,\xs_j, -i\xbs_j)$$
with $\vert\vert\Hb_P\vert\vert_2\leq 2M{\displaystyle{\frac{\varepsilon}{\delta}}}$ and~:
\be
\vert\vert H_*\vert\vert_{3/2}\leq\eta M\frac{\varepsilon}{\delta}\
\text{\it with}\ \eta =40\frac{M\pi}{\omega}\left(\frac{\rho}{\sigma} +\frac{\varepsilon}{\rho\sigma\delta}\right)
\label{taille_moyenne}
\ee

\end{theorem}
The proof of this theorem is given in section \ref{sec:app1}.

 The decreasing factor $\eta$ in the averaged perturbation can be minimized for a fixed lower bound on the mutual distance
$\delta >0$ and a mass ratio $\varepsilon$.
 We first relate the analyticity width $\rho$ to $\delta$ and $\varepsilon$ by choosing
$\rho ={\displaystyle{\sqrt{\frac{\varepsilon}{\delta}}}}$ such that the two terms in the factor $\eta$ are of the same order, then
$\eta ={\displaystyle{80\frac{M\pi}{\omega}\frac{\sqrt\eps}{\sigma\sqrt\delta}}}$.
 For $\delta\!\leq\! 16C\sigma_0$, the maximal analyticity width $\sigma ={\displaystyle{
 \frac{\delta}{32C}}}$ gives the minimal factor:
  $$
  \min (\eta) =2560\frac{CM\pi}{\omega}\sqrt{\frac{\eps}{\delta^3}}\ \text{for}\, \rho ={\displaystyle{\sqrt{\frac
    {\varepsilon}{\delta}}}}\ \text{and} \, \sigma = \frac{\delta}{32C}
  $$
which imposes a lower bound $\delta\geq\mathcal{O}(\eps^{1/3})$ in order to get a decreased perturbation in the averaged
system.
 Hence we recover the size of the Hill region inside which the averaged heliocentric Hamiltonian is not close to the initial one.

 \section{The averaged Hamiltonian and its topology}
\label{sec:averadgedro}
\subsection{Lagrange and Euler configurations in the averaged system}
\label{sec:EuLa}
Euler and Lagrange configurations are central configurations of the three body problem\footnote{See the webpages by A. Chenciner (2012) and by R. Moeckel (2014) at \\
www.scholarpedia.org/article/Three\_body\_problem, and at \\
 www.scholarpedia.org/article/Central\_configurations\#Euler.
}.
The two Lagrange configurations correspond to the situation where  the three bodies occupy the vertices of an equilateral triangle (the equilibrium points $L_4$ and $L_5$ in the circular RTBP), while the  three Euler's ones are the aligned configurations ($L_1$, $L_2$ and $L_3$ in the circular RTBP).  In terms of heliocentric elliptic elements, these are represented by two homothetical ellipses. The motion of the planets on these  fixed ellipses is Keplerian and  the difference of their mean longitude $\zeta_1 = \lam_1 - \lam_2$ is constant.
More precisely, in the Lagrange's case, we have:
\be
 \zeta_1 = \lam_1 -\lam_2 = \varpi_1 - \varpi_2 = \pm\pi/3,\,  e_1 = e_2,\,    a_1 = a_2,
\ee
while Euler configurations lead to the relations
\be
 \zeta_1 = \lam_1 -\lam_2 = \varpi_1 - \varpi_2 = 0, \,  e_1 = e_2,\,    a_1 =  a_2 + \gO(\eps^{1/3}) ,
\ee
if the two planets are on the same side of the Sun ($L_1$, $L_2$),  and to
\be
 \zeta_1 = \lam_1 -\lam_2 = \varpi_1 - \varpi_2 = \pi, \,  e_1 = e_2,\,    a_1 =  a_2 + \gO(\eps) ,
\ee
if the Sun is between the planets ($L_3$).
In the both cases, these trajectories, in fixed reference frame, are periodic orbits whose period is the common mean motion of the planets. As a result, these central configurations correspond to fixed points of the averaged problem.
More precisely, each type of configuration ($L_1$ to $L_5$) defines, in the averaged problem, a one-parameter family of equilibrium points.  This implies that a given fixed point of one of these families possesses an eigenvector (tangent to the family) associated to a zero eigenvalue.  This is the source of the degeneracy that  will be discussed in section \ref{sec:Hquad}.

\subsection{Some properties of the averaged Hamiltonian  }
\label{sec:propri}

In this section we will study the main properties of the averaged Hamiltonian of order $1$:
\be
\begin{split}
\Hb(\zetas_1, \Zs_j,\xs_j, -i\xbs_j)     &   = H'(\zetas_j,\Zs_j, \xs_j, -i\xbs_j) - H_*(\zetas_j, \Zs_j, \xs_j, -i\xbs_j)  \\
                                                 &  =  H_K(\Zs_1,\Zs_2)\! +\!\Hb_P(\zetas_1,\Zs_j,\xs_j, -i\xbs_j)
\end{split}
\label{eq:ham_moy_1}
\ee
and of its associated dynamics.

The  Hamiltonian $\Hb_P$ being an analytic function on the domain $\mathcal{K}_{\!{\frac{3}{2}}}$, it can be expanded in Taylor series in a  neighborhood of $(\xs_1,\xs_2) = (0,0)$ as:
\be
\Hb_P(\zetas_1, \Zs_j, \xs_j, -i\xbs_j)    =   \sum_{(\bfp, \bar\bfp)\in \mathbb{N}^4} C_{\bfp, \bfq}(\zetas_1,\Zs_1,\Zs_2)\xs_1^{p_1}\xs_2^{p_2}\xbs_1^{\bar p_1}\xbs_2^{\bar p_2}.
\label{eq:Taylor}
\ee
 In the previous summation, only the coefficients $C_{\bfp, \bfq}$ satisfying the relation 
%The planetary Hamiltonian being invariant by rotation, the  integers occurring in the previous summation satisfy the relation
\be
 p_1 +   p_2 =  \bar p_1 +  \bar p_2
 \label{eq:D'Al}
\ee
are different from zero.
 This propriety, known as D'Alembert rule, is equivalent to the fact that the quantity
 \be
 \vert \xs_1\vert^2 + \vert \xs_2\vert^2
 \ee
 is an integral of the averaged motion. The existence of this constant of motion makes possible the reduction of the averaged problem, leading to a Hamiltonian system which depends on two angles: the difference of the mean longitudes and the difference of the longitudes of the perihelion \citep[see][]{GiuBeMiFe2010}. This reduction, which decreases the number of degrees of freedom of the averaged problem from $3$ to $2$,  introduces some technical issues (addition of a parameter, singularity when the eccentricities tend to zero). For this reason, we prefer not to reduce the problem.

 Besides the reduction, the relation (\ref{eq:D'Al}) has many implications on the dynamics of the averaged system.

 One of these properties lies in the fact that the manifold
 \be
 \cC_0^{\mathbb{C}} = \{  (\zetas_j, \Zs_j, \xs_1, \xs_2, \xbs_1, \xbs_2) \in \mathcal{K}_{\!{\frac{3}{2}}} , \,
 \text{such that }\,             \xs_j = \xbs_j  = 0 \}
 \ee
  is an invariant manifold of the averaged Hamiltonian (\ref{eq:ham_moy_1}).  Indeed, the relation  (\ref{eq:D'Al}) implies that the Taylor series  (\ref{eq:Taylor}) starts at degree two. As a consequence
 \be
 \partial_{\xs_j}\Hb_P(\zetas_1, \Zs_j, 0,0) =  \partial_{\xbs_j}\Hb_P(\zetas_1, \Zs_j, 0,0) = 0,
 \ee
  which proves the invariance of the manifold $\cC_0^{\mathbb{C}}$  by the flow of (\ref{eq:ham_moy_1}).

   Actually, with the previous definition, $\cC_0^{\mathbb{C}}$ is a complex manifold and we will consider the real manifold \be
   \cC_0 =\cC_0^{\mathbb{C}}\cap (]\Delta , 2\pi -\Delta [\times\mathbb{R}\times\mathbb{C}
 \times\mathbb{C})\times (\mathbb{T}\times\mathbb{R}\times\mathbb{C}\times\mathbb{C})
 \ee
   which is also invariant by the flow of (\ref{eq:ham_moy_1}) since it is a real Hamiltonian.

\medskip

\subsection{The invariant manifold $\cC_0$}
\subsubsection{{Hamiltonian dynamics on $\cC_0$}}
The dynamics on $\cC_0$ is given by the restriction of the averaged Hamiltonian (\ref{eq:ham_moy_1}) to this manifold, that is:
\be
 \Hb_0'(\zetas_1, \Zs_1,\Zs_2) = H_K(\Zs_1,\Zs_2)  +\!\Hb_P(\zetas_1,\Zs_1,\Zs_2, 0,0,0,0)
\label{eq:dyn_C_0}
\ee

\be
\Hb_0' =   \Hb_0  + \Hb_0^{*},
\ee
with
\be
\Hb_0 =  -\frac{\beta_1\mu_1}{2a_1} - \frac{\beta_2\mu_2}{2a_2}
 +\eps \cG m_1m_2
\left(
\dfrac{\cos\zetas_1}{\sqrt{a_1a_2}}  - \dfrac{1}{\sqrt{a_1^2 + a_2^2 - 2a_1a_2\cos\zetas_1}}
\right)
\label{eq:def_Hb_0}
\ee
and
\be
\Hb_0^* =   \eps \frac{\cG m_1m_2}{\sqrt{a_1a_2}}
\left(
\frac{\beta_1\beta_2}{m_1m_2}\frac{ \sqrt{\mu_1\mu_2} }{\cG m_0} - 1
\right)
\cos\zetas_1 .
\ee

In the expression (\ref{eq:def_Hb_0}), $a_j$ can be easily expressed in term of $\Zs_j$  using the expression:
\be
a_1 = \mu_1^{-1}\beta_1^{-2}\left(
\Lam_1^0  + \Zs_1
\right)^2 ,
\quad
a_2 = \mu_2^{-1}\beta_2^{-2}\left(
\Lam_2^0  + \Zs_2 - \Zs_1
\right)^2
\ee
deduced from (\ref{eq:transf_lin}).

 With the expressions $\beta_j = m_j(1+\gO(\eps ))$ (resp. $\mu_j =G m_0 +\gO(\eps )$) and
 the analyticity of the functions $f(x,y)=x.y$ (resp. $g(x,y)=\sqrt{x.y}$) over $\R^2$ (resp.
 $\R^*\times\R^*$), we can write
 $$\left\vert\frac{\beta_1\beta_2}{m_1m_2}\frac{ \sqrt{\mu_1\mu_2} }{\cG m_0} - 1\right\vert < M\eps$$
where $M>0$ is independent of the small parameters in the problem.

 Moreover, we have a uniform upper bound on $\left\vert\frac{\cos\zetas_1}{\sqrt{a_1 a_2}}\right\vert$ (or equivalently on $\left\vert\frac{\cos\zetas_1}{\Lambda_1\Lambda_2}\right\vert$) over the compact
$\mathcal{K}_{\!\frac{3}{2}}$ and gathering these estimates yield:
 \be
\left\vert\left\vert \Hb_0^*\right\vert\right\vert_{\!{\frac{3}{2}}}< M\eps^2 .
\ee

\subsubsection{{Phase portrait  on $\cC_0$}}

\begin{figure}[h]
\begin{center}
   \begin{tikzpicture}
\path (0,0) node {\includegraphics[width=11cm]{\figpath/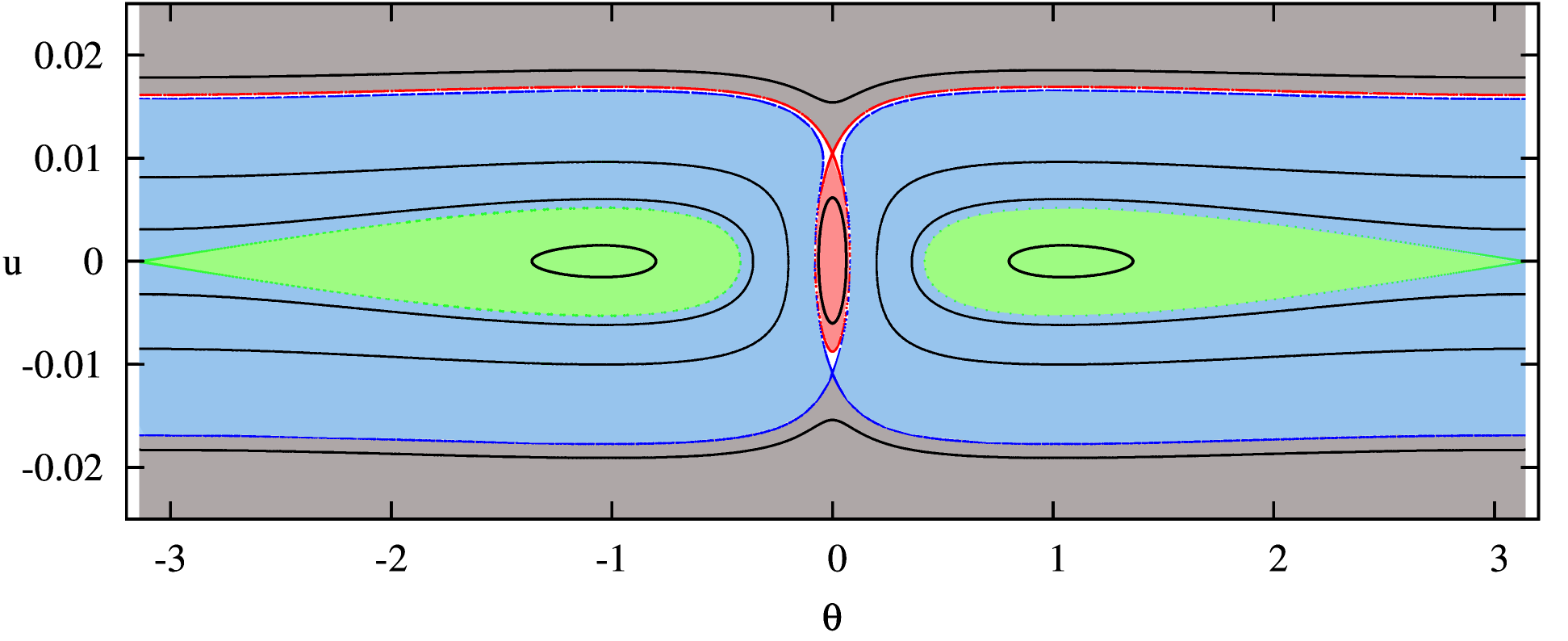}};
%
%%%%%%   VOIR CAHIER 14/04/2015 pour u en fonction de Z_1   %%%%%%%%%%%%%
 \draw[fill,white] (0,-2.2) -- (0,-1.9) -- (1,-1.9) -- (1,-2.2) -- cycle;
\path (0.4,-2.15) node  { $\zetas_1$ };
  \end{tikzpicture}
\caption{  Phase portrait of the Hamiltonian $\Hb_0$ in the coordinates $(\zetas_1,u)$.  The units and the parameter are chosen  such that $\Zs_2 = 0$, $\cG=1$, $\omega = 2\pi$, $\eps m_1 = 10^{-3}$,  $\eps m_2 = 3\times10^{-4}$.  See the text for more details.
 }
\label{fig:H_0}
\end{center}
\end{figure}

The phase portrait of the Hamiltonian $\Hb_0$ is represented on Fig. \ref{fig:H_0} in the plan $(\zetas_1,u)$ where $u$ is a dimensionless quantity  related to the action $\Zs_1$ and to the semi-major axes by
\be
u = \frac{2\omega^{1/3}\mu_0^{-2/3}}{m_1+m_2}\Zs_1  = \frac{m_1\sqrt{a_1} - m_2\sqrt{a_2}}{2}\sqrt{\mu_0} + \gO(\eps) \,
\ee
with $\mu_0 = \cG m_0$.
This figure has been obtained for particular values of the masses and of the parameters given in the  caption of the  figure \ref{fig:H_0}, but the qualitative structure of the phase space does not depend on these values.

The phase portrait of $\Hb_0$ is invariant by the symmetry with respect to the $\Zs_1 $axis  $(\Zs_1,\zetas_1) \longmapsto (\Zs_1,2\pi - \zetas_1)$. When the two planetary masses are equal, the Hamiltonian $\Hb_0$, and consequently its phase portrait, is  also invariant by the symmetry   $(\Zs_1,\zetas_1) \longmapsto (-\Zs_1, \zetas_1)$.

The shaded areas indicate the outside of the co-orbital resonance. In the upper grey region where $\Zs_1>0$, the angle $\zetas_1$ circulates clockwise while its circulation is anti-clockwise in the lower grey region ($\Zs_1<0$). The others domains correspond to the different kind of resonant motions.

The two elliptic fixed points in the middle of two green areas, located at
\be
 \zetas_1 \in\{\pi/3, 5\pi/3\} \qtext{and}  \Zs_1  = \frac{\eps}{6}\frac{m_1 - m_2}{m_1+m_2}\frac{m_1m_2}{m_0}\mu_0^{2/3}\omega^{-1/3} +\gO(\eps^2)\, ,
\ee
 are associated to the Lagrange equilateral configurations (see section \ref{sec:EuLa}). These points correspond to the relative equilibria where  the three bodies occupy the vertices of an equilateral triangle rotation at the constant angular velocity $\omega$. Each of these points, named $L_4$ and $L_5$ in the RTBP, is surrounded by tadpole orbits (green regions)  corresponding to periodic deformations of the equilateral triangle. These regions, whose maximal width in the $\Zs_1$-direction is of order $\eps^{1/2}$ \citep[see][for more details]{RoPo2013} are bounded by the separatrix $\cS_3$ that originates from the hyperbolic fixed point $L_3$ at
 \be
 \zetas_1 = \pi,  \quad \Zs_1  = \frac{\eps}{2}\frac{m_1 - m_2}{m_1+m_2}\frac{m_1m_2}{m_0}\mu_0^{2/3}\omega^{-1/3} +\gO(\eps^2)\, ,
 \ee
  for which the three bodies are aligned and the Sun is between the two planets and its separatrix. Outside this curve, in the blue domain, one find the horseshoe orbits that surround the three fixed points mentioned above.
Along these orbits, the angle $\zetas_1$  undergoes large variation such that $\zetas_1 \in [\zeta_m, 2\pi -\zeta_m]$  with $0<\gO(\eps^{1/3}) <\zeta_m <2\arcsin((\sqrt2 -1)/2) + \gO(\eps)$.

The horseshoe region (in blue), whose vertical extend is of order $\eps^{1/3}$,  is enclosed by the separatrices associated to  the two last Euler configurations located at $\zetas_1 = 0, \Zs_1 = \pm \gO(\eps^{1/3})$.  These points, corresponding to  the equilibria point  $L_1$ en $L_2$ in the RTBP, are associated with the Euler configurations for which the two planets are on the same side of the Sun.

 The last domain (in red), centred at the singularity of $\Hb_0$, that is the collision between the two planets, is surrounded by the  separatrix connecting the fixed point located at $\zetas_1 =0$ and $\Zs_1>0$ to itself. Inside this small region,  the two planets seem to be subjected to a prograde satellite-like motion, the one revolving the other one clockwise.
This last region (in red) and its neighbourhood that includes $L_1$ and $L_2$ are located at a distance to the collision of order $\eps^{1/3}$ and consequently, is outside the validity domain of the resonant normal form (see Theorem \ref{theo2} in the section \ref{sec:PerturbTheory}) . Indeed, in this region the remainder $H_*$ is at least of the same order that the perturbation $\Hb_P$.

Finally, let us remark that this figure is similar to the well known Hill's diagram (or zero-velocity curves) of the non averaged planar circular RTBP \cite[see][]{Sze1967} although the zero-velocity curves are not orbits  of the system. It is also topologically equivalent to the phase space of the averaged planar circular RTBP when the eccentricity of the massless body is equal to zero \citep{NeThoFeMo02, Morbidelli02}.

%%%

%%%%%%%%%%%%%%%%%%%%%%%%%%%%

\section{An integrable approximation of the averaged Hamiltonian}
\label{sec:simplified}

\subsection{Expansion around the resonance}
%

%%%%%%%%%

  In order to get a more tractable expression of the averaged Hamiltonian $\Hb_0$ in the domaine  ${\cal K}_{\!\frac32}$, it will be expanded in the neighborhood of $\Zs_j =0$ so that the remainder is always smaller than $H_P$. We will prove that the last condition is fulfilled  when the expansion of the  Keplerian part is truncated at the second order and $\Hb$ at zero order.

\subsubsection{Approximation of the Keplerian part}
\label{sec:approxKepler}

{
Let us start with the Keplerian part of the Hamiltonian. If its constant part $
2^{-1}\left(  \beta_1 \mu_1^{2/3} +   \beta_2 \mu_2^{2/3} \right)\omega^{2/3}
$
is omitted, $H_K$ can be written as:
\be
H_K(\Zs_1,\Zs_2)  =    \omega \Zs_2  + Q(\Zs_1,\Zs_2) +  R_1(\Zs_1,\Zs_2) + R_2(\Zs_1,\Zs_2)
\ee
\be
\text{with} \quad
\left\{
\begin{array}{ll}
 R_1(\Zs_1,\Zs_2)  & = \widetilde Q(\Zs_1,\Zs_2)  - Q(\Zs_1,\Zs_2) \\
 & \\
 R_2(\Zs_1,\Zs_2) & = H_K(\Zs_1,\Zs_2)  - \omega \Zs_2 -\widetilde Q(\Zs_1,\Zs_2)
\end{array}
\right.
\label{eq:R_12}
\ee
where the quadratic form $\widetilde Q$ reads:
\be
\begin{disarray}{rl}
\widetilde Q(\Zs_1,\Zs_2) =   &-\frac32\omega^{4/3} ( \beta_1^{-1} \mu_1^{-2/3} + \beta_2^{-1} \mu_2^{-2/3}) \Zs_1^2 \\
& + 3\omega^{4/3} \beta_2^{-1} \mu_2^{-2/3} \Zs_1\Zs_2
-\frac32\omega^{4/3}  \beta_2^{-1} \mu_2^{-2/3} \Zs_2^2,
\end{disarray}
\ee
%%%
and its approximation $Q(\Zs_1,\Zs_2)$:
\be
\label{quad}
\begin{disarray}{rl}
Q(\Zs_1,\Zs_2) =
  &-\frac32\omega^{4/3} \mu_0^{-2/3}( m_1^{-1} + m_2^{-1} ) \Zs_1^2 \\
& +3\omega^{4/3} \mu_0^{-2/3} m_2^{-1} \Zs_1\Zs_2
  -\frac32\omega^{4/3} \mu_0^{-2/3} m_2^{-1}  \Zs_2^2 ,
\end{disarray}
 \ee
 }

 {
 We first look at the  terms $R_1$, as:
  $$\beta_j = m_j + \gO(\eps)\ {\text{\rm and}}\ \mu_j = \cG m_0 + \gO(\eps) = \mu_0 +\gO(\eps),$$
  the quantity $R_1 = \widetilde Q - Q$ satisfies the relation:
  \be
 R_1  = \gO(\eps\vert\vert (\Zs_1 ,\Zs_2)\vert\vert^2)
 \ee
and more specifically, there exists a large enough constant $M>0$ independent of $\varepsilon$, $\rho$, $\sigma$ and $\delta$
such that
$$\vert\vert R_1\vert\vert_{\!{{3}\over{2}}}\leq M\varepsilon\rho^2.$$
}

{
 As regards the estimate of $R_2$, the application of the Taylor formula on the function $g(t)=H_K(t\Zs_1,t\Zs_2)$ for $(\Zs_1,\Zs_2)\in\mathcal{K}_{\!{{3}\over{2}}}$ leads to:
$$
R_2(\Zs_1,\Zs_2) = H_K(\Zs_1,\Zs_2) - \omega \Zs_2 - {\widetilde{Q}}(\Zs_1,\Zs_2) =\int_0^1\frac{(1-t)^2}{2}g^{(3)}(t) dt \,
$$
 Using the inequality (\ref{eq:born_der}), we have for all $t\in [0,1]$:
 $\vert g^{(3)}(t)\vert\leq M\vert\vert (\Zs_1,\Zs_2)\vert\vert^3$, which  finally leads to:
\be
R_2(\Zs_1,\Zs_2)=\gO_3 (\Zs_1,\Zs_2),
\ee
and more specifically to: $\vert\vert R_2\vert\vert_{\!{{3}\over{2}}}\leq{\displaystyle{\frac{9M}{16}}}\rho^3$.}

\subsubsection{Approximation of the perturbation}
\label{sec:approxpert}

{We consider the function $G(\zetas_1,\xs_j,\xbs_j)$ on the compact $\mathcal{K}_{\!{{3}\over{2}}}$ which is
equal to the modified function $\Hb_P(\zetas_1,0,0,\xs_j, \xbs_j)$ where the resonant actions $\Lam_j^0$ equal to $\mu_j^{2/3}\beta_j\omega^{-1/3}$ are replaced by $\mu_0^{2/3} \beta_j\omega^{-1/3}$.
 Equivalently, using the notation:
$$
\Delta Z_j^0 = \sqrt[6]{\frac{\mu_0}{\mu_j}}\sqrt[3]{\frac{\mu_j^2}{\omega}}\beta_j - \sqrt[3]{\frac{\mu_j^2}{\omega}}\beta_j  =\gO
(\varepsilon )\ {\text{\rm for}}\ j\in\{ 1,2\}
$$
and using the transformation (\ref{eq:transf_lin}) we obtain:
\be
G(\zetas_1,\xs_j,\xbs_j) =\Hb_P(\zetas_1,\Delta Z_1^0,\Delta Z_1^0 +\Delta Z_2^0,\xs_j, \xbs_j).
\ee

 Then the averaged perturbation $\Hb_P$ can be split in the sum of three term as follows:
\be
\Hb_P (\zetas_1,\Zs_1,\Zs_2,\xs_j,\xbs_j)  =G(\zetas_1,\xs_j,\xbs_j) + R_3(\zetas_1,\xs_j, \xbs_j) + R_4(\zetas_1,\Zs_j, \xs_j, \xbs_j)
\ee
with
\be
\left\{
\begin{array}{ll}
\ \ \ R_3(\zetas_1,\xs_j, \xbs_j)  & =\Hb_P(\zetas_1,0,0,\xs_j, \xbs_j) -G(\zetas_1,\xs_j,\xbs_j)\\
 & \\
 R_4(\zetas_1,\Zs_j, \xs_j, \xbs_j) & = \Hb_P (\zetas_1,\Zs_1,\Zs_2,\xs_j,\xbs_j)   -\Hb_P(\zetas_1,0,0,\xs_j, \xbs_j)
 \end{array}
\right.
\label{eq:R_34}
\ee
}

{ In order to estimate theses remainders, we need to consider the smaller compact
 $\mathcal{K}_{\!{{3}\over{2}}}^{(\kappa )}\subset\mathcal{K}_{\!{{3}\over{2}}}$ where the
 neglected terms are smaller than the averaged perturbation $\vert\vert \Hb_P\vert\vert^{
 (\kappa )}_{\!{{3}\over{2}}}\leq\vert\vert H_P\vert\vert_{\!{{3}\over{2}}}$.
}

 With similar reasonings as in the previous section, we use the mean value theorem to evaluate the
remainder in the truncation at order 0 of $\Hb_P$ we consider $\mathcal{K}_p^{(\kappa )}
\subset\mathcal{K}_p$ which is defined for $p>0$ and $\kappa >0$ by:
$$\mathcal{K}_p^{(\kappa )} :=\! =\!\left\{\left(\zeta_j,Z_j,\xi_j ,\eta_j\right)_{j\in\{ 1,2\} }\!
 \in\mathcal{K}_p\ {\text{\rm such that}}\ \max (\vert Z_1\vert ,
 \vert Z_2\vert )\leq\kappa\rho\right\} .$$
 The supremum norm $\vert\vert .\vert\vert_{\infty}$ on the space of holomorphic functions over
the compact $\mathcal{K}^{(\kappa )}_p$ will be denoted by $\vert\vert .\vert\vert^{(\kappa )}_p$.
 For $p=3/2$, we obtain on this smaller compact $\mathcal{K}^{(\kappa )}_{\!{{3}\over{2}}}
 \subset\mathcal{K}_{\!{{3}\over{2}}}$ for a small enough $\kappa >0$:
$$\vert\vert R_3\vert\vert^{(\kappa )}_{\!{{3}\over{2}}}=\vert\vert\Hb_P(\zetas_1,0,0,\xs_j, \xbs_j) - \Hb_P
(\zetas_1,\Delta Z_1^0,\Delta Z_1^0 +\Delta Z_2^0,\xs_j, \xbs_j)\vert\vert^{(\kappa )}_{\!{{3}\over{2}}}$$
%$$\Longrightarrow\vert\vert R_3\vert\vert_{\!{{3}\over{2}}}^{(\kappa )}\leq\vert\vert\partial_Z\Hb_P\vert
%\vert_{\!{{3}\over{2}}}\vert\vert (\Delta Z_1^0,\Delta Z_1^0 +\Delta Z_2^0)\vert\vert\leq\frac{2}{\rho}
%\vert\vert\Hb_P\vert\vert_2\vert\vert (\Delta Z_1^0,\Delta Z_1^0 +\Delta Z_2^0)\vert\vert
%$$
which implies that
$$
\vert\vert R_3\vert\vert_{\!{{3}\over{2}}}^{(\kappa )}\leq\vert\vert\partial_Z\Hb_P\vert
\vert_{\!{{3}\over{2}}}\vert\vert (\Delta Z_1^0,\Delta Z_1^0 +\Delta Z_2^0)\vert\vert\leq\frac{2}{\rho}
\vert\vert\Hb_P\vert\vert_2\vert\vert (\Delta Z_1^0,\Delta Z_1^0 +\Delta Z_2^0)\vert\vert
$$
We have an upper bound $\vert\vert (\Delta Z_1^0,\Delta Z_1^0 +\Delta Z_2^0)\vert\vert\leq C\varepsilon$
for some constant $C>0$ and our estimate on $H_P$ yields:
{
\be
\vert\vert\Hb_P\vert\vert_2\leq 2M\frac{\varepsilon}{\delta}\Longrightarrow\vert\vert  R_3\vert\vert^{(\kappa )}_{\!{{3}\over{2}}}\leq M\frac{\varepsilon^2}{\rho\delta}
\ee
for a large enough constant $M>0$.
}

 In the same way, the mean value theorem yields:
{
$$
\vert\vert R_4\vert\vert^{(\kappa )}_{\!{{3}
\over{2}}}\leq\vert\vert\partial_Z\Hb_P\vert\vert_{\!{{3}\over{2}}}\vert\vert (\Zs_1 ,\Zs_2)\vert\vert
\leq\frac{2}{\rho}
\vert\vert\Hb_P\vert\vert_2\vert\vert (\Zs_1 ,\Zs_2)\vert\vert
$$
}
{
\be
\Longrightarrow\vert\vert  R_4 (\zetas_1,\Zs_1,\Zs_2,\xs_j,\xbs_j) \vert\vert^{(\kappa )}_{\!{{3}
\over{2}}}\leq 4M\frac{\varepsilon}{\rho\delta}\kappa\rho =4M\frac{\varepsilon}{\delta}\kappa .
\ee
}

%
% Finally, we replace $\Lam_j^0 = \mu_j^{2/3}\beta_j\omega^{-1/3}$ by $\mu_0^{2/3}m_j\omega^{-1/3}$ in $\Hb_P(\zetas_1,0,
% 0,\xs_j, \xbs_j)$, consequently with the estimate $\vert\vert\Hb_P\vert\vert_2$ :
%\be
% \label{eq:G}
% \Hb_P(\zetas_1,\Zs_1,\Zs_2,\xs_j, \xbs_j) =  G(\zetas_1,\xs_j, \xbs_j) + G_*(\zetas_1,\Zs_1,\Zs_2,\xs_j, \xbs_j)
%\ee
%where $ G(\zetas_1,\xs_j, \xbs_j)  =  \Hb_P(\zetas_1,0,0,\xs_j, \xbs_j)\vert_{\Lam_j^0 = \mu_j^{2/3}\beta_j\omega^{-1/3}}$
%and
%\be
%\vert\vert G_*\vert\vert^{(\kappa )}_{\!{{3}\over{2}}}\leq 4M\frac{\varepsilon}{\delta} (\kappa +\gO(\varepsilon )).
%\ee

\subsubsection{The final approximation of $\Hb$}

 Gathering the approximation given in sections \ref{sec:approxKepler} and \ref{sec:approxpert}, and omitting the constant
 term $2^{-1}\left(  \beta_1 \mu_1^{2/3} +   \beta_2 \mu_2^{2/3} \right)\omega^{2/3}$, the averaged Hamiltonian $\Hb$ takes
 the following form:
\be
\Hb(\zetas_j\Zs_j,\xs_j,\xbs_j ) = \omega \Zs_2 + Q(\Zs_1,\Zs_2)  +  G(\zetas_1,\xs_j, \xbs_j) + R(\zetas_j\Zs_j,\xs_j,
\xbs_j ) ,
\label{eq:approxHb}
\ee
{where the quadratic part $Q$ is given by the expression (\ref{quad}), the perturbation $G$ is defined in the top of Section \ref{sec:approxpert},  and the remainder $R$ is defined by the sum $R = R_1 + R_2 + R_3 + R_4$ where  the $R_j$ are given in (\ref{eq:R_12}) and (\ref{eq:R_34}).}

  If, as in the end of the section \ref{sec:PerturbTheory}, we relate the analyticity width $\rho$ to $\delta$ and $\varepsilon$ by  $\rho^2\delta =\eps$, we get the following upper bound with $M$ large enough:
\be
\vert\vert R\vert\vert^{(\kappa )}_{\!{{3}\over{2}}}\leq M\rho\left(\varepsilon\rho +\frac{9}{16}\rho^2 +\varepsilon +
4\kappa\rho\right)\leq M\rho\left( 3\rho^2 +4\kappa\rho\right)
\ee
since $\rho <1$ and $\delta <1\Longrightarrow\varepsilon <\rho^2$.

 Especially, we have $\vert\vert R\vert\vert^{(\kappa )}_{\!{{3}\over{2}}}<3M\rho^3 +4M\kappa\rho^2 <m\rho^2
<\vert\vert \Hb_P\vert\vert_{\!{{3}\over{2}}}\leq\vert\vert \Hb_P\vert\vert^{(\kappa )}_{\!{{3}\over{2}}}$ for
$\kappa$ small enough and $\varepsilon$ small enough (since $\rho <\varepsilon$).

\medskip

\subsection{The dynamics on $\cC_0$ and its implication for the initial problem}
\label{sec:sol_C_0}
Using the approximation (\ref{eq:approxHb}) of the averaged Hamiltonian where the remainder $R(\zetas_j\Zs_j,\xs_j,\xbs_j )$ is neglected, its restriction to the invariant manifold $\cC_0$ reads:
\be
\widetilde\cH_0 = \omega \Zs_2 + Q(\Zs_1,\Zs_2) + \eps\mu_0^{2/3}\omega^{2/3}\frac{ m_1m_2}{m_0} F(\zetas_1)
\label{eq:sol_simpl_C0}\ee
with
\be
F(\zetas_1) = G(\zetas_1,0,0) = \cos\zetas_1 - \frac{1}{\sqrt{2 - 2\cos\zetas_1}}
\ee
In order to uncouple the fast and semi-fast degrees of freedom, we define the symplectic linear map $(\zetas_1,
\zetas_2, \Zs_1,\Zs_2) =\cL (\varphi_1,\varphi_2,I_1,I_2)$ on the cylinder $]\Delta ,2\pi -\Delta [\times\T\times\R\times\R$  by:
\be
\bpm
\zetas_1 \\
\zetas_2
\epm
=
\bpm
1 & 0 \\
-\frac{m_1}{m_1+m_2} & 1
\epm
\bpm
\varphi_1 \\
\varphi_2
\epm
\quad ,
\bpm
\Zs_1 \\
\Zs_2
\epm
=
\bpm
1 & \frac{m_1}{m_1+m_2} \\
 0 & 1
\epm
\bpm
I_1 \\
I_2
\epm
\label{eq:zeta_phi}
\ee
and completed by the identity in the $\xs_j$ and $\xbs_j$ variables.
As a consequence, we have:
\be
\begin{split}
\cH_0(\varphi_j,I_j) & = \widetilde\cH_0 \circ \cL(\varphi_j,I_j)  =   \cH_0^{(1)}(\varphi_1,I_1)  + \cH_0^{(2)}(\varphi_2,I_2) \\
 & = -\frac32\omega^{4/3} \mu_0^{-2/3}\left(\frac{1}{m_1} + \frac{1}{m_2}  \right) I_1^2
+ \eps\mu_0^{2/3}\omega^{2/3}\frac{ m_1m_2}{m_0} F(\varphi_1) \\
 & +  \omega I_2 - \frac32\omega^{4/3} \mu_0^{-2/3}\frac{I_2^2}{m_1+m_2} \\
\end{split}
\label{eq:H_simple}
\ee

The dynamics of the  fast variables $(\varphi_2,I_2)$ is now  governed by $\cH_0^{(2)}$  while the dynamics of the semi-fast variables $(\varphi_1,I_1)$ is given by  $\cH_0^{(1)}$. Of course, these two Hamiltonians are integrable. But the dynamics of $\cH_0^{(2)}$  is trivial while that of $\cH_0^{(1)}$ is less.

In the domain $ {\cal K}_{\!\frac32}^{(\kappa )}$, the phase portrait of the two approximations $\cH_0^{(1)}$ \citep[already explored by several authors, i.e.][and references therein]{Morais2001} and $\Hb_0$ of the averaged Hamiltonian are topologically equivalent. They both have  two elliptic fixed points corresponding to $L_4$ and $L_5$ and an unstable equilibrium associated to the Euler configuration $L_3$. The stable equilibria are located at $(\varphi_1,I_1) = (\pm\pi/3, 0)$, and their eigenvalues, which are the same for both points, are equal to  $\pm i\sqrt{\eps}\sqrt{\frac{27}{4}\frac{m_1+m_2}{m_0}}\omega + \gO(\eps^{3/2})$. The unstable point is located at $(\varphi_1,I_1) = (\pi, 0)$, its eigenvalues are $\pm \sqrt{\eps}\sqrt{\frac{21}{8}\frac{m_1+m_2}{m_0}}\omega + \gO(\eps^{3/2})$.
As we have seen in the section \ref{sec:propri}, in the domain enclosed by the separatrices emanating from this fixed point, one find tadpole orbits surrounding $L_4$ or $L_5$. Outside these invariant manifolds are the horseshoe orbits that encompass the three fixed points mentioned above.

{   At this points, we know at least qualitatively what are the orbits on the invariant manifold $\cC_0$. To go further, we would like to have the temporal parametrization of the corresponding trajectories. However, even if the Hamiltonian $\cH_0^{(1)}$ is integrable, its trajectories cannot be given explicitly. Consequently, in the sequel, we will assume that these trajectories that satisfy the canonical differential equations:
%\be
%\left\{
%  \begin{split}
% \dot I_1 & = \eps\mu_0^{2/3}\omega^{2/3}\frac{m_1m_2}{m_0}\left( 1 - \left(2 - 2\cos\varphi_1\right)^{-3/2}\right) \sin\varphi_1 \\
% %
% \dot\varphi_1  &= -3\frac{m_1+m_2}{m_1m_2}\omega^{4/3}\mu_0^{-2/3}I_1
% \end{split}
% \right.
% \label{eq:diif_syst}
%\ee
\be
\left\{
  \begin{array}{ll}
 \dot I_1 & = \eps\mu_0^{2/3}\omega^{2/3}\dfrac{m_1m_2}{m_0}\left( 1 - \left(2 - 2\cos\varphi_1\right)^{-3/2}\right) \sin\varphi_1 \\
 \dot\varphi_1  &= -3\dfrac{m_1+m_2}{m_1m_2}\omega^{4/3}\mu_0^{-2/3}I_1
 \end{array}
 \right.
 \label{eq:diif_syst}
\ee

are perfectly known, once given its initial conditions $(\varphi_1(0), I_1(0))$.
 Actually, these solutions are all periodic since the level curves of $\mathcal{H}_0^{(1)}$
 are closed and without singularities.
}
%

%%%%%%%%%%%%%%%%%%%%%%%%%%%%
\subsection{The normal stability of the manifold $\cC_0$}
\label{sec:Hquad}

 Now, we study the linearized dynamic around the invariant
manifold $\mathcal{C}_0$.
%
% Consider an arbitrary trajectory $(\varphi_1(t), I_1(t))$ on $\cC_0$, we
% have the variational equation:
%\be
%\begin{split}
%\Hb_2(\xs_j,\xbs_j) =  \eps\omega \frac{m_1 m_2}{m_0} \Bigg(
%& \frac{A(\zetas_1)}{m_1}\xs_1\xbs_1
% +  \frac{B(\zetas_1)}{\sqrt{m_1m_2}}\xs_1\xbs_2  \\
%%
%& + \frac{\Bb(\zetas_1)}{\sqrt{m_1m_2}}\xbs_1\xs_2
% + \frac{A(\zetas_1)}{m_2}\xs_2\xbs_2
% \Bigg)
% \end{split}
%\label{eq:Hquadh}
%\ee
%%
%with
%%
%\be
%\begin{split}
% A(\zetas_1) &  = \frac{1}{4 D(\zetas_1)^5}\left(
%   5\cos2\zetas_1 - 13 + 8\cos\zetas_1
%\right)
% - \cos\zetas_1, \\
%%
% B(\zetas_1)  &=  e^{-2i\zetas_1}      -\frac{1}{8 D(\zetas_1)^5}
% \left({\displaystyle
%  e^{-3i\zetas_1}  + 16e^{-2i\zetas_1} - 26e^{-i\zetas_1}   + 9 e^{i\zetas_1}
%} \right), \\
%%
% D(\zetas_1) & = \sqrt{2 - 2\cos\zetas_1}.
% \end{split}
% \label{eq:coeffHquad}
%\ee
%\bigskip
%The variational equation in the $(\xs_j,\xbs_j)$ direction around a given produce solution $\zetas_1(t)$ of the equation (\ref{eq:sol_simpl_C0}) reads:
%%
%\be
%\dot X = M(t) X
%\label{eq:variations}
%\ee
%where
%\be
%X =
%\begin{pmatrix}
%x_1\\
%x_2
%\end{pmatrix}
%%
%%
%\qtext{and }
%M(t) =  i\eps \omega \frac{m_1m_2}{m_0}
%\begin{pmatrix}
%\dfrac{A(\zetas_1(t))}{m_1} & \dfrac{\Bb(\zetas_1(t))}{\sqrt{m_1m_2}} \\
%\dfrac{B(\zetas_1(t))}{\sqrt{m_1m_2}} & \dfrac{A(\zetas_1(t))}{m_2}
%\end{pmatrix}
%\label{eq:matrix_floquet}
%\ee
%
%%%%%%%%%%%%%%%%%%%
Consider an arbitrary trajectory  on $\cC_0$. It is shown in \cite{RoPo2013} that  the variational equation in the $(\xs_j,\xbs_j)$ direction around this solution  reads:
\be
\dot X = M(t) X
\label{eq:variations}
\ee
where
\be
X =
\begin{pmatrix}
x_1\\
x_2
\end{pmatrix}
\qtext{and }
M(t) =  i\eps \omega \frac{m_1m_2}{m_0}
\begin{pmatrix}
\dfrac{A(\zetas_1(t))}{m_1} & \dfrac{\Bb(\zetas_1(t))}{\sqrt{m_1m_2}} \\
\dfrac{B(\zetas_1(t))}{\sqrt{m_1m_2}} & \dfrac{A(\zetas_1(t))}{m_2}
\end{pmatrix}
\label{eq:matrix_floquet}
\ee
with
\be
\begin{split}
 A(\zetas_1) &  = \frac{1}{4 D(\zetas_1)^5}\left(
   5\cos2\zetas_1 - 13 + 8\cos\zetas_1
\right)
 - \cos\zetas_1, \\
 B(\zetas_1)  &=  e^{-2i\zetas_1}      -\frac{1}{8 D(\zetas_1)^5}
 \left({\displaystyle
  e^{-3i\zetas_1}  + 16e^{-2i\zetas_1} - 26e^{-i\zetas_1}   + 9 e^{i\zetas_1}
} \right), \\
 D(\zetas_1) & = \sqrt{2 - 2\cos\zetas_1},
 \end{split}
 \label{eq:coeffHquad}
\ee
$\zetas_1(t)$ being a solution of the canonical equation associated to the Hamiltonian (\ref{eq:sol_simpl_C0})

%\bigskip
 According to the Floquet theorem \citep[see][]{MeHa1992}, if the frequency of the considered periodic solution is $\nu$, the solutions of the variational equation take the form
\be
Y(t) = P(\nu t) \exp(U t),
\ee
where $U$ is a constant matrix and $P(\psi)$ is a matrix whose coefficients are $2\pi$-periodic functions of $\psi$.
As, if $Y$ is a fundamental matrix solution to the variational equation along a $2\pi/\nu$-periodic solution,  one has the relation
\be
Y(t + 2\pi\nu^{-1}) = Y(t)\exp\left(
2\pi\nu^{-1}U\right).
\ee
It turns out that the stability of the solutions of the variational equation (\ref{eq:variations}) depends on the eigenvalues of the matrix $U$. As stated in section \ref{sec:propri}, the quantity $\xs_1\xbs_1 +\xs_2\xbs_2$ is an integral of the variational equation  (\ref{eq:variations}). This implies that the solutions of (\ref{eq:variations}) are bounded, and as a consequence, $U$ is diagonalisable and the real parts of its eigenvalues are equal to zero.  Although we cannot exclude that one of the eigenvalue vanishes, the invariant manifold $\cC_0$ is normally stable.

There exists, on $\cC_0$, at least  three trajectories for which one eigenvalue of  the variational equation (\ref{eq:variations}) vanishes.
These are the stationary solutions of the differential system (\ref{eq:diif_syst}), that is the Euler configuration $L_3$ and the Lagrange ones $L_4$ and $L_5$.

For the collinear configuration  associated to $L_3$, which corresponds to $(\varphi_1, I_1) = (\pi,0)$, we have
\be
A(\pi) =  \frac78, \quad B(\pi) = \frac78.
\ee
As a consequence, the equilibrium has  two eigendirections collinear to
\be
V_{\pi}^1 =
\begin{pmatrix}
\sqrt{m_2} \\
\sqrt{m_1}
\end{pmatrix}
\qtext{and}
V_{\pi}^2 =
\begin{pmatrix}
\sqrt{m_1} \\
-\sqrt{m_2}
\end{pmatrix}
\label{eq:vectp}
\ee
associated respectively to the eigenvalue
\be
v_{\pi}^1 = i\eps\omega\ \frac78\frac{m_1+m_2}{m_0}  \qtext{ and} v_{\pi}^2 = 0.
\ee
The reason why one of the eigenvalues vanishes has been given in section \ref{sec:EuLa}. Indeed it is easy to verify that  the eigenvector $V_{\pi}^2$ corresponds to  the elliptic Euler's configurations where the two planet are in the two sides of the Sun.

The other eigendirection corresponds to a non trivial family of periodic orbits.  According to the expressions (\ref{eq:vectp}), the configurations corresponding to $V_{\pi}^1$ are two ellipses in conjonction $(\varpi_1 = \varpi_2)$ with equal semi-major axis and whose eccentricities satisfy the relation $m_1 e_1 = m_2e_2$.

Contrary to the previous Euler's configurations, the ellipses are not fixed, but precess at the same rate defined by the frequency $-i v_{\pi}^1 = \gO(\eps)$.   This eigendirection gives rise to a one-parameter family unstable periodic orbits of the averaged problem (periodic in rotating frame in the non-averaged problem) described by \cite{HaPsyVo2009}  and related to the Poincar\'e solutions of second sort  \cite[see][for more details]{RoPo2013}.

Similar phenomena occur in the neighborhood of the two equilateral fixed points $L_4$ and $L_5$ at $\varphi_1 = \pm\pi/3$ and $I_1 = 0$.
Without entering into details \citep[see][]{RoPo2013}, let us just mention what we get for  $L_4$ (the results are similar for the other equilateral equilibrium). The coefficients of the matrix (\ref{eq:matrix_floquet}) satisfy
\be
A(\frac{\pi}{3}) = -\frac{27}{8}, \quad B(\frac{\pi}{3}) = \frac{27}{16}(1 - i\sqrt3).
\ee
As a consequence, its eigenvectors are
\be
V_{\pi/3}^1 =
\begin{pmatrix}
\sqrt{m_2}e^{i\pi/3} \\
-\sqrt{m_1}
\end{pmatrix}
\qtext{and}
V_{\pi/3}^2 =
\begin{pmatrix}
\sqrt{m_1}e^{i\pi/3} \\
\sqrt{m_2}
\end{pmatrix}
\label{eq:vectp_L4}
\ee

and the corresponding  eigenvalues are
\be
v_{\pi/3}^1 = -i\eps\omega\ \frac{27}{8}\frac{m_1+m_2}{m_0}  \qtext{ and} v_{\pi/3}^2 = 0.
\ee
It is easy to verify that the configurations associated to $V_{\pi/3}^2$  are the elliptic equilateral configurations that are fixed points of the averaged Hamiltonian. This is  the reason why $v_{\pi/3}^2 = 0$. As in the case of the Euler configuration, the eigenvector $V_{\pi/3}^1$ is tangent to a one-parameter family of periodic orbits, called anti-Lagrange by \cite{GiuBeMiFe2010}. For small eccentricities, the elliptic elements of the corresponding orbits that  precess  simultaneously at the frequency $iv_{\pi/3}^1$  satisfy the  relation $m_1e_1 = m_2 e_2$ and $\varpi_1 - \varpi_2 = \lam_1 - \lam_2  + \pi$.

%%%%%%%%%%%%%%%%%
\section{Consequences for the co-orbital motion }
\label{se:cons_coorb}

So far, the study of the manifold $\cC_0$ and its dynamics was carried out in the context of the averaged problem.
To end this section, we will study what happens to the dynamics of $\cC_0$ when we go back to the initial variables.
 The following theorem gives a partial approximation, on a finite but large time, of the dynamics on the manifold $\cC_0$
in the initial variables $(\lam_j,\Lam_j,x_j,-i\xb_j)$. Actually, we explain at the end of this section the way
to obtain a complete result of approximation with our methods.

\medskip

 We consider a solution in the initial variables $(\lam_j(t),\Lam_j(t),x_j(t),-i\xb_j(t))$ starting with at $t=0$ in the image of the manifold $\mathcal{C}_0$ by the transformation $(\lam_j,\Lam_j,x_j,-i\xb_j)
 =\Psi\circ\mathcal{C}\circ\mathcal{L}(\varphi_j,I_j, \xs_j, -i\xbs_j)$ considered in section 2, 3 and 5, hence:
 $$(\lam_j(0),\Lam_j(0),x_j(0),-i\xb_j(0))=\Psi\circ\mathcal{C}\circ\mathcal{L}(\varphi_j(0),I_j(0),0,0).$$

 Using these linear and averaging transformations, we can formally write:
\be
\begin{split}
\lam_1(t)  &= \widetilde\omega(I_2(0)) t + \frac{m_2}{m_1+m_2}\varphi_1(t) + \varphi_2(0) + \rho_1(t) \\
\lam_2(t)  &= \widetilde\omega(I_2(0)) t - \frac{m_1}{m_1+m_2}\varphi_1(t) + \varphi_2(0) + \rho_2(t) \\
\Lam_1(t) &= \Lam_1^0  + I_1(t) + \frac{m_2}{m_1+m_2} I_2(0) + \tau_1(t) \\
\Lam_2(t) &=  \Lam_2^0  - I_1(t) + \frac{m_1}{m_1+m_2} I_2(0) + \tau_2(t) \\
x_j(t)        &= \tau_{j+2}(t)
\end{split}
\ee
where the function $(\varphi_1 (t),I_1 (t))$ in these expressions is the solution of the differential system (\ref{eq:diif_syst}) with the initial conditions given as follow:
\be
\begin{split}
\varphi_1(0)  & = \lam_1(0) - \lam_2(0) \\
I_1(0)  &= \frac{ m_2}{m_1+m_2}(\Lam_1(0) -\Lam_1^0) - \frac{ m_1}{m_1+m_2}(\Lam_2(0) -\Lam_2^0) \\
\varphi_2(0) & =   \frac{ m_1}{m_1+m_2}\lam_1(0)  + \frac{ m_2}{m_1+m_2}\lam_2(0)  \\
I_2(0) &=  \Lam_1(0) -\Lam_1^0 + \Lam_2(0) -\Lam_2^0 \\
\widetilde\omega(I_2)  &=  \partial_{I_2}\cH_0^{(2)}(\varphi_2,I_2)  =  \omega  - 3\omega^{4/3} \mu_0^{-2/3}\frac{I_2}{m_1+m_2}
\end{split}
\ee
and the remainders $\rho_1 (t)$, $\rho_2 (t)$, $\tau_1 (t)$, $\tau_2 (t)$, $\tau_{j+2} (t)$ should be small over large times.

 We give a partial theorem in this direction.

\medskip

 We first recall bounds on the remainders in the previous computations with the choices $\delta\rho^2=\varepsilon$ and $32C\sigma =\delta <16 C\sigma_0$, there exists a large enough constant $M$ independent of $\varepsilon$, $\rho$, $\sigma$ and $\delta$:
\be
\vert\vert H_*\vert\vert_{3/2}\leq M\sqrt{\frac{\eps^3}{\delta^5}}
\ee
\be
\left\vert\left\vert \Hb_0^*\right\vert\right\vert_{\!{{3}\over{2}}}=\eps \gO(\eps )=\gO(\eps^2)\leq M\eps^2
\ee
\be
\vert\vert R\vert\vert^{(\kappa )}_{\!{{3}\over{2}}}\leq M\rho\left( 3\rho^2 +4\kappa\rho\right)
\ee
and we denote
\be
\left\vert\left\vert H_* +\Hb_0^* + R\right\vert\right\vert^{(\kappa )}_{\!{{3}\over{2}}} < L:= M\left[\sqrt{\frac{\eps^3}{\delta^5}}+\eps^2 +\rho\left( 3\rho^2 +4\kappa\rho\right)\right] .
\ee

\medskip

\begin{theorem}
\label{theo3}
 We denote:
 $$p=\frac{m_1 +m_2}{2m_1 +m_2}\ {\text{\it and}}\ \mathcal{T}_p\ {\text{\it the first time of escape out of}}\  \mathcal{K}_p^{(\kappa )}$$
for the solution $\Phi_t^{H'\circ\mathcal{L}}(\varphi_j(0),I_j(0),0,0)$ along the flow governed by the averaged Hamiltonian $H'$.

 With the previous assumptions and notations, we have the following bounds on the remainders:
\be
\vert\tau_{j+2}(t)\vert\leq\rho\ {\text{\it for}}\ \vert t\vert\leq\min\left(\frac{\kappa\rho^2}{4L},\mathcal{T}_p\right)
\ee
and 
\be
\begin{split}
 \Lam_1(t) + \Lam_2(t) &= \Lam_1(0)+\Lam_2(0)+ \tau(t)\ 
 \\ &{\text{\it with}}\ \vert\tau(t)\vert\leq\rho\ {\text{\it for}}\ \vert t\vert\leq\min\left(\frac{\kappa\rho^2}{4L},\mathcal{T}_p\right) .
\end{split}
\ee

\end{theorem}

\medskip

The proof of this theorem is given in the section \ref{seq:theo_3} and exactly the same reasonings would give a complete approximation of the solutions on $\mathcal{C}_0$ except that we need moreover the action-angles variables linked to the integrable Hamiltonian $\cH_0^{(1)}$ which can be built by classical techniques (cf Arnold)

 The first step is to bound the difference between the flows linked to two nearby Hamiltonian in the normalized variables $(\varphi_j,I_j, \xs_j, -i\xbs_j)\in\mathcal{K}^{(\kappa )}_p$ and then gives a sharp timescale such that these two flows remain close in the initial variables $(\lam_j,\Lam_j,x_j,-i\xb_j)$.

\bigskip

%%%%%%%%%%%%%%%%%
\section{Proof of the theorems}
%%%
\subsection{Proof of theorem \ref{theo1}}
\label{sec:app1}

 Using the notations of section \ref{r11} and the mean value theorem, we obtain a lower bound for the minimal distance (with $\C^2$ equipped with the Hermitian
  norm) between two orbits in the complex domain $\mathcal{K}^{(\mathbb{C})}_{\Delta ,\rho_0 ,\sigma_0}$.

  Actually, denoting
  $$(\brt_1^{(\C)},\br_1^{(\C)},\brt_2^{(\C)},\br_2^{(\C)})=\Upsilon\left(\varsigma_j,z_j,\xi_j ,\eta_j\right)_{ j\in\{ 1,2\} }\ \text{\rm with}\ \Upsilon =(\Phi_1\circ\Psi ,\Phi_2\circ\Psi ),$$
we have:
  $$(\brt_1^{(\R)},\br_1^{(\R)},\brt_2^{(\R)},\br_2^{(\R)})=\Psi\left( (\text{\rm Re}(\varsigma_1),0,0,0);(\text{\rm Re}(\varsigma_2),0,0,0)\right) ,
  $$
 and the mean value theorem together with our bound $C$ on the differential of $\Upsilon$ allows to write for $j\in\{ 1,2\}$:
\begin{equation*}
\begin{split}
 &\left\vert\left\vert\left(\varsigma_j,z_j,\xi_j ,\eta_j\right) -(\text{\rm Re}(\varsigma_j),0,0,0)\right\vert\right\vert
  \leq\rho +2\sqrt{\rho\sigma}+\sigma\\
  &\Longrightarrow
  \left\vert\left\vert\br_j^{(\C)}-\br_j^{(\R)}\right\vert\right\vert\leq C(\rho +2\sqrt{\rho\sigma}+\sigma )
\end{split}
  \end{equation*}
then the triangular inequality yields:
  \be
  \begin{array}{ccc}
  \left\vert\left\vert\br_1^{(\C)}-\br_2^{(\C)}\right\vert\right\vert & = &\left\vert\left\vert\br_1^{(\C)}-\br_1^{(\R)}+
  \br_1^{(\R)}-\br_2^{(\R)}+\br_2^{(\R)}-\br_2^{(\C)}\right\vert\right\vert\hfill\\
  & \geq &\left\vert\left\vert\br_1^{(\R)}-\br_2^{(\R)}\right\vert\right\vert -\left\vert\left\vert\br_1^{(\C)}-
    \br_1^{(\R)}\right\vert\right\vert -\left\vert\left\vert\br_2^{(\C)}-\br_2^{(\R)}\right\vert\right\vert\hfill\\
  & \geq &\left\vert\left\vert\br_1^{(\R)}-\br_2^{(\R)}\right\vert\right\vert -2C(\rho +2\sqrt{\rho\sigma}+\sigma )\hfill\\
  \end{array}
 \label{eq:ineg_r_C}
  \ee

   If we assume that
   \be
   \rho <\sigma\ \text{\rm and}\ \sigma\leq \frac{\delta}{16C}=\frac{a}{8C}\sin\left(\frac{\Delta}{2}\right) ,
 \ee
   we obtain with the lower bound on the mutual distance in the real domain:
   $$
   \left\vert\left\vert\br_1^{(\C)}-\br_2^{(\C)}\right\vert\right\vert\geq\delta -8C\sigma\geq\frac{\delta}{2}=
   a\sin\left(\frac{\Delta}{2}\right)
   $$
   and the expression of the planetary Hamiltonian gives the upper bounds on the size of $H_P$ in the complex domain.

\medskip

   In the same way, under the assumption $\rho <\sigma <1$, we obtain with the upper bound on the mutual distance in the real domain:
   \be
   \left\vert\left\vert\br_1^{(\C)}-\br_2^{(\C)}\right\vert\right\vert\leq  M +8C\sigma < M +8C
   \ee
   and the expression of the planetary Hamiltonian gives the lower bounds on the size of $H_P$ in the complex domain provided that $M$ is large enough to satisfy ${\displaystyle{\frac{\varepsilon}{M}}}<M +8C$.

\medskip

\subsection{Proof of theorem \ref{theo2}}

 Let $\Delta >0$, $\rho >0$ and $\sigma >0$ such that $(\Delta ,2\rho ,2\sigma )$ satisfy (\ref{seuil_1}), hence:
   \be
   \rho <\sigma\leq \frac{\delta}{32C}=\frac{a}{16C}\sin\left(\frac{\Delta}{2}\right)\ \text{\rm with}\
   \delta :=2a\sin\left(\frac{\Delta}{2}\right)
 \ee
then there exists a constant $M>0$ independant of $\varepsilon$, $\Delta$, $\rho$ and $\sigma$ such that
\be
\vert\vert H_K\vert\vert_{2\rho}<M\ \text{\rm and}\ \vert\vert H_P\vert\vert_2 =\vert\vert H_P
\vert\vert_{\Delta , 2\rho ,2\sigma }<M\frac{\varepsilon}{\delta}\label{seuil_6}
\ee
for the supremum norm $\vert\vert .\vert\vert_{2\rho}$ (resp. $\vert\vert .\vert\vert_2 =\vert\vert .
\vert\vert_{\Delta ,2\rho ,2\sigma}$) on the space of holomorphic functions over the compact
$$\{ (Z_1,Z_2)\in\C^2\ \text{\rm such that}\ \max (\vert Z_1\vert ,\vert Z_2\vert )\leq 2\rho\}$$
 (resp. over the compact $\mathcal{K}^{(\mathbb{C})}_{\Delta ,2\rho ,2\sigma }=\mathcal{K}_2$).

 In view of our bounds (\ref{seuil_Pert}) and (\ref{seuil_6}), we obtain:
\be
\vert\vert\chi\vert\vert_2 <{\Frac{2\pi}{\omega}}\left( \vert\vert H_P\vert\vert_2\right)<M
\Frac{2\pi\varepsilon}{\omega\delta}\ .\label{seuil_Fonct_Gen}
\ee

 Moreover, with our rescaling of the masses, the derivatives of the Keplerian part $H_K$ remains of order one and
we can also assume that the constant $M>0$ independant of $\varepsilon$, $\Delta$, $\rho$ and $\sigma$ gives also an
upper bound on the Hessian of $H_K$~:
\be
\forall (Z_1,Z_2)\in\C^2\ \text{\rm with}\ \max (\vert Z_1\vert ,\vert Z_2\vert )\leq 2\rho\ :\
\left\vert\frac{\partial^2 H_K}{\partial Z_i\partial Z_j}(Z_1,Z_2)\right\vert <M.\label{seuil_Hess}
\ee

\medskip

 We must first estimate the size of the partial derivatives and the Poisson
bracket of analytical functions on $\mathcal{K}_2$ by classical applications of
Cauchy inequalities which can be found in \cite{Poschel1993} and  \cite{Giorgilli2003}.
%Pöschel, Jürgen  Nekhoroshev estimates for quasi-convex Hamiltonian systems.
%Math. Z. 213 (1993), no. 2, 187–216.
% Giorgilli, Antonio Exponential stability of Hamiltonian systems. Dynamical systems. Part I, 87–198,
%Publ. Cent. Ric. Mat. Ennio Giorgi, Scuola Norm. Sup., Pisa, 2003.
%%
%%%
% Giorgilli, A.: Exponential stability of Hamiltonian systems. In: Dynamical Sys-
%tems. Part I. Hamiltonian Systems and Celestial Mechanics. Selected papers
%from the Research Trimester held in Pisa, Italy, February 4 Ð April 26, 2002.
%Pubblicazioni del Centro di Ricerca Matematica Ennio de Giorgi, Proceedings.
%Scuola Normale Superiore, Pisa (2003), 87Ð198
\medskip

 Let $f$ be analytical on $\mathcal{K}_2$ (continuous on the boundary), we can write~:
$$
\left\vert\left\vert\Dron{f}{\zeta}(\zeta , Z, x, -i\xb)\right\vert \right\vert =
\Sup_{\vert\vert (e_1,e_2)\vert\vert =1} \left\vert \left\vert {\Der{}{t}}_{\vert
t=0} f(\zeta +te, Z, x, -i\xb )\right\vert \right\vert
$$
for $(\zeta , Z, x, -i\xb )=(\zeta_j , Z_j, x_j, -i\xb_j ) _{ j\in\{ 1,2\} }\in\mathcal{K}_2$.

\medskip

 One then applies Cauchy formula to the function $t\mapsto f(\zeta +te, Z, x, -i\xb )$
of the complex variable $t$, holomorphic and continuous on the boundary for $\vert t\vert\leq\sigma /4$ when
$(\zeta , Z, x, -i\xb )\in\mathcal{K}^{(\mathbb{C})}_{7/4}$, and obtains~:
$$\left\vert\left\vert\Dron{f}{\zeta}\right\vert \right\vert_{7/4}
\leq \Frac{4}{\sigma }\vert\vert f\vert\vert_2\ .$$

 The same reasoning yields the equivalent inequalities for the other partial
derivatives~:
$$\left\vert \left\vert \Dron{f}{Z} \right\vert \right\vert_{7/4}\leq
\Frac{4}{\rho}\vert\vert f\vert\vert_2\ ;
\ \left\vert \left\vert \Dron{f}{x}\right\vert\right\vert_{7/4}\ \!\!\text{\rm and}\ \left\vert
\left\vert \Dron{f}{\xb}\right\vert\right\vert_{7/4}\leq\Frac{4}{\sqrt{\rho\sigma}}\vert\vert f\vert\vert_2\ .$$

 To estimate the size of the Poisson brackets for two analytical functions on
${\mathcal K}_2$, we write in a similar way that
$$\{ f,g\} (\zeta , Z, x, -i\xb)\! =\! {\frac{d}{dt}
}_{\vert t=0}\! \left[\! g\! \left(\! \zeta\! -\! t {\frac{\partial f}{\partial Z}}
 , Z\! +\! t {\frac{\partial f}{\partial \zeta}} , x\! -\! it {\frac{\partial f}
{\partial\xb }} , -i\xb\! +\! t {\frac{\partial f}{\partial x}}\right)\right]$$
and the function of the complex variable $t$ in the
right hand side is defined for $\vert t\vert\leq\frac{\rho\sigma}{16\vert\vert f\vert\vert_2}$
(we apply the Cauchy formula to $f$), hence we can write:
\begin{equation*}
\begin{split}
& \vert\vert\{ f,g\}\vert\vert_{7/4}\leq 16\Frac{\vert\vert f\vert\vert_2\vert\vert g\vert\vert_2}
{\rho\sigma},\ {\text{\rm in the same way we obtain }}\\
&\vert\vert\{ f,g\}\vert\vert_{3/2}\leq 4\Frac{\vert\vert f\vert\vert_2\vert\vert g\vert\vert_2}{\rho\sigma}.
\end{split}
\end{equation*}

\medskip

 Since the averaging transformation $\varphi$ is the time-one map $\Phi_{1}^{\chi}$ of the Hamiltonian flow
generated by some auxiliary hamiltonian $\chi$, for any function $K$ defined on the phase space:
%$$\frac{d}{dt}(K\circ\Phi_{t}^{\chi})=\{ K,\chi\}\circ\Phi_{t}^{\chi}\Longrightarrow\zetas_j -\zeta_j =\int_0^1\frac
%{\partial\chi}{\partial Z_j}\circ\Phi_{t}^{\chi}dt\Longrightarrow\left\vert\left\vert\zetas_j -\zeta_j\right\vert
%\right\vert_{3/2}\leq\left\vert\left\vert\frac{\partial\chi}{\partial Z_j}\right\vert\right\vert_{7/4}$$
%
%
\begin{equation*}
\begin{split}
\frac{d}{dt}(K\circ\Phi_{t}^{\chi})=\{ K,\chi\}\circ\Phi_{t}^{\chi} & \Longrightarrow\zetas_j -\zeta_j =\int_0^1\frac
{\partial\chi}{\partial Z_j}\circ\Phi_{t}^{\chi}dt \\
& \Longrightarrow\left\vert\left\vert\zetas_j -\zeta_j\right\vert
\right\vert_{3/2}\leq\left\vert\left\vert\frac{\partial\chi}{\partial Z_j}\right\vert\right\vert_{7/4}\end{split}
\end{equation*}
for $j\in\{ 1,2\}$. Actually, we prove that starting inside $\mathcal{K}_{3/2}$ along the flow $\Phi_{1}^{\chi}$ yields
a time of escape out of $\mathcal{K}_{7/4}$ which is bigger than 1 and we can use our estimates on $\chi$.

Hence , we have:
$$\left\vert\left\vert\zetas_j -\zeta_j\right\vert\right\vert_{7/4}\leq
\Frac{4}{\rho}\vert\vert\chi\vert\vert_2 <\Frac{8M\pi}{\omega}\Frac{\varepsilon}{\rho\delta}<\Frac{\sigma}{4}$$
with our threshold (\ref{seuil_3}).

 In the same way, we obtain:
 $$\left\vert\left\vert\Zs_j -Z_j\right\vert\right\vert_{3/2}\leq\left\vert\left\vert\frac{\partial\chi}{\partial\zeta_j}
 \right\vert\right\vert_{7/4}\leq\Frac{4}{\sigma}\vert\vert\chi\vert\vert_2 <\Frac{8M\pi}{\omega}\Frac{\varepsilon}
 {\sigma\delta}<\Frac{\rho}{4}$$
 $$\left\vert\left\vert\xs_j -x_j\right\vert\right\vert_{3/2}\leq\left\vert\left\vert\frac{\partial\chi}{\partial
 \xb_j} \right\vert\right\vert_{7/4}\leq\Frac{4}{\sqrt{\rho\sigma}}\vert\vert\chi\vert\vert_2 <\Frac{8M\pi}{\omega}
 \Frac{\varepsilon}{\sqrt{\rho\sigma}\delta}<\Frac{\sqrt{\rho\sigma}}{4}$$
 $$\left\vert\left\vert\xbs_j -\xb_j\right\vert\right\vert_{3/2}\leq\left\vert\left\vert\frac{\partial\chi}{\partial
 x_j} \right\vert\right\vert_{7/4}\leq\Frac{4}{\sqrt{\rho\sigma}}\vert\vert\chi\vert\vert_2 <\Frac{8M\pi}{\omega}
 \Frac{\varepsilon}{\sqrt{\rho\sigma}\delta}<\Frac{\sqrt{\rho\sigma}}{4}$$
which yields ${\cal K}_{\!\frac54}\subseteq{\cal C}({\cal K}_{\!\frac32})\subseteq{\cal K}_{\!\frac74}$.

\medskip

\begin{equation*}
\begin{split}
\text{As } \; &H'(\zetas_j,\Zs_j, \xs_j, -i\xbs_j) =  H_K(\Zs_1,\Zs_2) \\
       &+ \Hb_P(\zetas_1,\Zs_j,\xs_j, -i\xbs_j) +  H_*(\zetas_j,\Zs_j,\xs_j, -i\xbs_j),
\end{split}
\end{equation*}
where
$$
H_*\! =\!\left(\nabla H_K\! -\! {\overrightarrow{\omega}}\right).\Dron{\chi}{\zetas}+\{ \chi ,H_P \} +
H' -H -\{ \chi ,H\}
\ \text{\rm for}\ {\overrightarrow{\omega}}\! :=\!\left(
                                   \begin{array}{c}
                                      0\\
                                     \omega\\
                                   \end{array}
                                 \right),$$
 we may use Taylor's formula at order two to write:
 $$H' -H -\{ \chi ,H\}=H\circ\Phi_{1}^{\chi}-H\circ\Phi_{0}^{\chi}-\frac{d}{dt}(H\circ\Phi_{0}^{\chi})
 =\int_0^1 (1-t)\{ \chi ,\{ \chi ,H\}\}\circ\Phi_{t}^{\chi} dt .$$

 Since $\Phi_{1}^{\chi}({\cal K}_{3/2} )\subset {\cal K}_{7/4}$, using the upper bound $M$ on the
Hessian of $H_K$ and Cauchy inequalities, one finds the following estimate~:
\be
\left\vert\left\vert\left (\nabla H_K\! -\! {\overrightarrow{\omega}}\right).\Dron{\chi}
{\zetas}\right\vert\right\vert_{7/4}\leq M\Frac{7}{4}\rho\Frac{\vert\vert\chi\vert\vert_2}{\sigma /4}\leq
\Frac{7\pi M^2}{\omega}\Frac{\rho\varepsilon}{\sigma\delta}
\ee
moreover~:
\be
\vert \vert\{\chi , H_P\} \vert \vert_{7/4}<16\Frac{\vert\vert
\chi\vert\vert_2\vert\vert H_P\vert\vert_2}{\rho\sigma}\leq\Frac{16\pi M^2}{\omega}\Frac{\varepsilon^2}{\rho
\sigma\delta^2} .
\ee

\medskip

 To estimate the third term in $H_*$, we insert again the definition of $\chi$ given
in section (\ref{equ_homo}) into the Poisson bracket to get~:
$$\{ \chi ,H\} =\Hb_P- H_P +\left(\nabla H_K - {\overrightarrow{\omega}}\right).\Dron
{\chi}{\zetas}+\{ \chi ,H_P \}$$
hence
%$$\vert\vert\{\chi ,\{\chi ,H\}\}\vert\vert_{3/2}\leq\vert\vert\{\chi ,H_P -\Hb_P\}\}\vert
%\vert_{3/2}\! +\left\vert\left\vert\left\{\chi ,\left(\nabla H_K - {\overrightarrow
%{\omega}}\right).\Dron{\chi}{\zetas}\right\}\right\vert\right\vert_{3/2}\! +\vert\vert
%\{\chi ,\{\chi ,H_P\}\}\vert\vert_{3/2}$$
%%
\begin{equation*}
\begin{split}
\vert\vert\{\chi ,\{\chi ,H\}\}\vert\vert_{3/2} & \leq\vert\vert\{\chi ,H_P -\Hb_P\}\}\vert
\vert_{3/2}\!  \\
&+\left\vert\left\vert\left\{\chi ,\left(\nabla H_K - {\overrightarrow
{\omega}}\right).\Dron{\chi}{\zetas}\right\}\right\vert\right\vert_{3/2}\! +\vert\vert
\{\chi ,\{\chi ,H_P\}\}\vert\vert_{3/2}
\end{split}
\end{equation*}

 Using again Cauchy inequalities, we can estimate the previous terms and the sum of these inequalities
yields the given value of $\eta$, indeed:
$$\vert\vert\{\chi ,H_P -\Hb_P\}\}\vert
\vert_{3/2}\! <4\Frac{\vert\vert\chi\vert\vert_2\vert\vert H_P-\Hb_P\vert\vert_2}{\rho\sigma}\leq\Frac{16\pi M^2}
{\omega}\Frac{\varepsilon^2}{\rho\sigma\delta^2} .$$
%$$
%\left\vert\left\vert\left\{\chi ,\left(\nabla H_K - {\overrightarrow{\omega}}\right).\Dron{\chi}{\zetas}\right\}
%\right\vert\right\vert_{3/2}\! <8\Frac{\vert\vert\chi\vert\vert_2\left\vert\left\vert\left(\nabla H_K -
%{\overrightarrow{\omega}}\right).\Dron{\chi}{\zetas}\right\vert\right\vert_{7/4}}{\rho\sigma}\leq 7M
%\left(\Frac{4\pi M}{\omega}\Frac{\rho\varepsilon}{\sigma\delta}\right)^2
%$$
%
\begin{equation*}
\begin{split}
\left\vert\left\vert\left\{\chi ,\left(\nabla H_K - {\overrightarrow{\omega}}\right).\Dron{\chi}{\zetas}\right\}
\right\vert\right\vert_{3/2}\! &  <8\Frac{\vert\vert\chi\vert\vert_2\left\vert\left\vert\left(\nabla H_K -
{\overrightarrow{\omega}}\right).\Dron{\chi}{\zetas}\right\vert\right\vert_{7/4}}{\rho\sigma} \\
 &\leq 7M
\left(\Frac{4\pi M}{\omega}\Frac{\rho\varepsilon}{\sigma\delta}\right)^2
\end{split}
\end{equation*}
$$\vert\vert\{\chi ,\{\chi ,H_P\}\}\vert\vert_{3/2} <8\Frac{\vert\vert\chi\vert\vert_2\vert\vert\{\chi ,H_P\}
\vert\vert_{3/4}}{\rho\sigma}\leq M\left(\Frac{16\pi M}{\omega}\Frac{\varepsilon}{\rho\sigma\delta}\right)^2
\Frac{\varepsilon}{\delta}$$
moreover the threshold $\Frac{\varepsilon}{\delta\rho\sigma}<\Frac{\omega}{32\pi M}$ gives:
\be
\Frac{\varepsilon}{\delta\sigma}<\Frac{\omega}{32\pi M}\rho\Longrightarrow\left\vert\left\vert\left\{\chi ,
\left(\nabla H_K - {\overrightarrow{\omega}}\right).\Dron{\chi}{\zetas}\right\}\right\vert\right\vert_{3/2}\!
<\Frac{7}{2}\Frac{\pi M^2}{\omega}\Frac{\rho}{\sigma}\Frac{\varepsilon}{\delta}
\ee
\be
\Frac{\varepsilon}{\delta\rho\sigma}<\Frac{\omega}{32\pi M}\Longrightarrow\vert\vert\{\chi ,\{\chi ,H_P\}\}
\vert\vert_{3/2} <\Frac{8\pi M^2}{\omega}\Frac{\varepsilon}{\rho\sigma\delta}\Frac{\varepsilon}{\delta}
\ee

\bigskip

Finally, the sum of all the previous estimates yields:
$$\vert\vert H_*\vert\vert_{3/2}\leq\frac{M\pi}{\omega}\left(\frac{21}{2}\frac{\rho}{\sigma} +
40\frac{\varepsilon}{\rho\sigma\delta}\right) M\frac{\varepsilon}{\delta}$$
and ensures the claimed value for the upper bound $\eta =40\Frac{M\pi}{\omega}\left(\Frac{\rho}{\sigma} +
\Frac{\varepsilon}{\rho\sigma\delta}\right)$.

%%%%

\subsection{ Proof of the theorem \ref{theo3}}
\label{seq:theo_3}

 The first part of the proof of theorem \ref{theo3} comes from the use of Cauchy inequalities to bound the size of the Hamiltonian vector field linked to $(H_* +\Hb_0^* + R)\circ\mathcal{L}$ over the compact $\mathcal{K}^{(\kappa )}_p$.

  The second part comes from the size of $\Psi\circ\mathcal{C}\circ\mathcal{L}$ composed of the averaging transformation $\mathcal{C}$ and two linear transformations $\Psi$ and $\mathcal{L}$ and the choice of a time $T$ which give terms of the same order in the upper bound on the error in the approximate solution.

  More specifically, by straightforward computations the norm of the transformation $\mathcal{L}$ admits the upper bound $p^{-1}=\frac{2m_1+m_2}{m_1+m_2}$ hence
  $$\mathcal{L}\left(\mathcal{K}^{(\kappa )}_p\right)\subset\mathcal{K}^{(\kappa )}_1 .$$

  With our notations, the function
$$(\varphi_1(t),\varphi_2(t), I_1(t), I_2(t),0,0)=(\varphi_1(t),\varphi_2(0) +\partial_{I_2}\cH_0^{(2)}(I_2(0)) t,I_1(t), I_2(0),0,0)$$
\be
(\varphi_1(t),\varphi_2(0) +\left(\omega  - 3\omega^{4/3} \mu_0^{-2/3}\frac{I_2(0)}{m_1+m_2}\right) t,I_1(t), I_2(0),0,0)
\ee
is a solution of the Hamiltonian system linked to $\cH_0^{(2)}$ on $\mathcal{C}_0$, hence it is a solution of the averaged system governed by $\Hb$.

 Then, the complete averaged Hamiltonian $H'$ satisfies $H'-\Hb = H_* +\Hb_0^* + R$ and
 if we denote $({\widetilde{\varphi}}_1(t),{\widetilde{\varphi}}_2(t), {\widetilde{I}}_1(t), {\widetilde{I}}_2(t),\xs (t),\xbs (t))$ the solution of the system linked to $H'$ starting at $(\varphi_1(0),\varphi_2(0), I_1(0), I_2(0),0,0)$,
 we can write
 $$\frac{d{\widetilde{I}}_2}{dt}=-\partial_{\varphi_2}(H_* +\Hb_0^* + R)\circ\mathcal{L}.$$

 Then, with our assumption on the time of escape $\mathcal{T}_p$, we have:
 $$\mathcal{L}({\widetilde{\varphi}}_1(t),{\widetilde{\varphi}}_2(t), {\widetilde{I}}_1(t), {\widetilde{I}}_2(t),\xs (t),\xbs (t))\in\mathcal{K}^{(\kappa )}_1\ {\text{\rm for}}\ \vert t\vert <\mathcal{T}_p$$
and Cauchy inequality yields:
\be
\left\vert\left\vert\frac{d{\widetilde{I}}_2}{dt}\right\vert\right\vert <\frac{2}{\kappa\rho}\left\vert\left\vert H_* +\Hb_0^* + R\right\vert\right\vert^{(\kappa )}_{\!{{3}\over{2}}}<\frac{2L}{\kappa\rho}.
\ee
$$\Longrightarrow {\widetilde{I}}_2(t)= {\widetilde{I}}_2(0)+ \tau_2(t)\ {\text{\it with}}\ \vert\tau_2(t)\vert\leq\frac{2L}{\kappa\rho}\vert t\vert\ {\text{\it for}}\ \vert t\vert\leq \mathcal{T}_p .$$

\medskip

 We have the same bound in the averaged variable $\Zs_2 =I_2$.

\medskip

 Finally, in the initial variables $Z_2$, we have:
 $$\left\vert\left\vert\Zs_2 -Z_2\right\vert\right\vert_{3/2}\leq\left\vert\left\vert\frac{\partial\chi}{\partial\zeta_2}
 \right\vert\right\vert_{7/4}<\Frac{8M\pi}{\omega}\Frac{\varepsilon}
 {\sigma\delta}<\Frac{\rho}{4}$$
et $\left\vert\left\vert Z_2 (t) -Z_2 (0)\right\vert\right\vert\leq\left\vert\left\vert Z_2 (t) -\Zs_2 (t)\right\vert\right\vert +\left\vert\left\vert\Zs_2 (t) -\Zs_2 (0)\right\vert\right\vert +\left\vert\left\vert\Zs_2 (0) -Z_2 (0)\right\vert\right\vert$
$$
\Longrightarrow\left\vert\left\vert Z_2 (t) -Z_2 (0)\right\vert\right\vert\leq
2\left\vert\left\vert\Zs_2 -Z_2\right\vert\right\vert_{3/2}+\left\vert\left\vert{\widetilde{I}}_2 (t) -{\widetilde{I}}_2 (0)\right\vert\right\vert
$$
\be
\Longrightarrow\left\vert\left\vert Z_2 (t) -Z_2 (0)\right\vert\right\vert\leq
\Frac{\rho}{2}+\frac{2L}{\kappa\rho}\vert t\vert <\rho\ {\text{\it for}}\ \vert t\vert\leq\min\left(\frac{\kappa\rho^2}{4L},\mathcal{T}_p\right)
\ee

which gives the formula for $\Lambda$ in the theorem.

 The formula for $(x_j,\xb_j )$ is proved in the same way.

%%%%%%%%%%%%%%%%%
%\section{Conclusion}

%%%%%%%%%%%%%%%%%
%
%%%%%\bibliographystyle{unsrt}
%
%\bibliographystyle{apalike}
%\bibliography{AR_RONI2015a}

%%%%%%%%%%%%%%%%

%%%%%%%%%%%%%%%%%

\end{document}